\documentclass[11pt]{article}
\pdfoutput=1
\usepackage{color}
\usepackage[textwidth = 430 pt, textheight = 630 pt]{geometry}
\definecolor{MyDarkBlue}{rgb}{0.15,0.25,0.45}
\usepackage{epsfig,rotating}
\usepackage{amsmath,amssymb}
\usepackage{amsfonts}
\usepackage{mathrsfs}
\usepackage{bbm}
\usepackage{bm}
\usepackage{hyperref}
\hypersetup{
colorlinks=true,
citecolor=MyDarkBlue,
linkcolor=MyDarkBlue,
urlcolor=MyDarkBlue,
pdfauthor={Sam Palmer and Christian S\"amann},
pdftitle={Constructing Generalized Self-Dual Strings},
pdfsubject={hep-th}
breaklinks=true
}

\let\fn\footnote
\renewcommand{\footnote}[1]{\linespread{1.1}\fn{#1}\linespread{1.29}}

\makeatletter\renewcommand{\section}{\@startsection
{section}{1}{\z@}{-3.5ex plus -1ex minus
    -.2ex}{2.3ex plus .2ex}{\bf }}
\makeatletter\renewcommand{\subsection}{\@startsection{subsection}{2}{\z@}{-3.25ex
plus -1ex minus
   -.2ex}{1.5ex plus .2ex}{\it }}
\makeatletter\renewcommand{\subsubsection}{\@startsection{subsubsection}{3}{-2.45ex}{-3.25ex
plus -1ex minus -.2ex}{1.5ex plus .2ex}{\it }}
\renewcommand{\thesection}{\arabic{section}}
\renewcommand{\thesubsection}{\arabic{section}.\arabic{subsection}}
\renewcommand{\@seccntformat}[1]{\@nameuse{the#1}.~~}

\renewcommand{\theequation}{\thesection.\arabic{equation}}
\makeatletter \@addtoreset{equation}{section}

\renewenvironment{thebibliography}[1]
     {\baselineskip=16pt plus 2pt minus 1pt
      \section*{\large\refname
        \@mkboth{\MakeUppercase\refname}{\MakeUppercase\refname}}%
     \list{\@biblabel{\@arabic\c@enumiv}}%
           {\settowidth\labelwidth{\@biblabel{#1}}%
            \leftmargin\labelwidth
            \advance\leftmargin\labelsep
            \@openbib@code
            \usecounter{enumiv}%
            \let\p@enumiv\@empty
            \renewcommand\theenumiv{\@arabic\c@enumiv}}%
      \sloppy
      \clubpenalty4000
      \@clubpenalty \clubpenalty
      \widowpenalty4000%
      \sfcode`\.\@m}

\newcommand{\appendices}{
\section*{Appendix}\label{appendices}\setcounter{subsection}{0}
\addcontentsline{toc}{section}{Appendix}
\setcounter{equation}{0}
\makeatletter
\renewcommand{\theequation}{\Alph{subsection}.\arabic{equation}}
\renewcommand{\thesubsection}{\Alph{subsection}}
\@addtoreset{equation}{subsection}
\makeatother
}

\renewcommand{\slash}[1]{#1\hspace{-0.27cm}/\,}

\def\periodb#1{\setbox0=\hbox{$#1$}#1\hskip-\wd0\hbox to\wd0{-}}
\newcommand{\nablas}{\slash{\nabla}}
\newcommand{\nablabs}{\slash{\bar{\nabla}}}

\newcommand{\binomr}[2]{\binom{\,#1\,}{\,#2\,}}

\newcommand{\xd}{\dot{x}}

\newcommand{\unit}{\mathbbm{1}}   			
\newcommand{\id}{\mathrm{id}}   			

\newcommand{\CA}{\mathcal{A}}    			

\newcommand{\CCC}{\mathscr{C}}

\newcommand{\CI}{\mathcal{I}}

\newcommand{\CL}{\mathcal{L}}

\newcommand{\CN}{\mathcal{N}}

\newcommand{\CT}{\mathcal{T}}

\newcommand{\frg}{\mathfrak{g}}				

\newcommand{\FR}{\mathbbm{R}}     			
\newcommand{\FC}{\mathbbm{C}}     			
\newcommand{\RZ}{\mathbbm{Z}}     			

\newcommand{\dd}{\mathrm{d}}     			
\newcommand{\dpar}{\partial}     			
\newcommand{\diag}{{\mathrm{diag}}}     		
\newcommand{\de}{\mathrm{e}}     			
\newcommand{\di}{\mathrm{i}}     			
\newcommand{\eps}{{\varepsilon}}			

\newcommand{\zb}{{\bar{z}}}
\newcommand{\Zb}{{\bar{Z}}}
\newcommand{\alphab}{{\bar{\alpha}}}
\newcommand{\betab}{{\bar{\beta}}}
\newcommand{\gammab}{{\bar{\gamma}}}

\newcommand{\psib}{{\bar{\psi}}}

\newcommand{\sigmab}{{\bar{\sigma}}}

\newcommand{\eand}{{~~~\mbox{and}~~~}}     		
\newcommand{\ewith}{{~~~\mbox{with}~~~}}

\newcommand{\der}[1]{\frac{\dpar}{\dpar #1}}   		
\newcommand{\dder}[1]{\frac{\dd}{\dd #1}}   		
\newcommand{\delder}[1]{\frac{\delta}{\delta #1}}   		
\newcommand{\tr}{\,\mathrm{tr}\,}     			

\newcommand{\au}{\mathfrak{u}}
\newcommand{\asu}{\mathfrak{su}}

\newcommand{\sU}{\mathsf{U}}     			

\newcommand{\sSU}{\mathsf{SU}}

\newcommand{\sSO}{\mathsf{SO}}

\newcommand{\sSpin}{\mathsf{Spin}}

\newcommand{\acton}{\vartriangleright}     			
\newcommand{\remark}[1]{}     				
\def\tyng(#1){\hbox{\tiny$\yng(#1)$}}			
\def\tyoung(#1){\hbox{\tiny$\young(#1)$}}			

\newcommand{\dotsp}{\;\cdot\;}

\begin{document}
\begin{titlepage}
\begin{flushright}
 HWM--11--12 \\ EMPG--11--13
\end{flushright}
\vskip 2.0cm
\begin{center}
{\LARGE \bf Constructing Generalized Self-Dual Strings}
\vskip 1.5cm
{\Large Sam Palmer and Christian S\"amann}
\setcounter{footnote}{0}
\renewcommand{\thefootnote}{\arabic{thefootnote}}
\vskip 1cm
{\em Department of Mathematics\\
Heriot-Watt University\\
Colin Maclaurin Building, Riccarton, Edinburgh EH14 4AS, U.K.\\
and Maxwell Institute for Mathematical Sciences, Edinburgh,
U.K.}\\[0.5cm]
{Email: {\ttfamily sap2@hw.ac.uk~,~c.saemann@hw.ac.uk}}
\end{center}
\vskip 1.0cm
\begin{center}
{\bf Abstract}
\end{center}
\begin{quote}
We generalize a recently developed ADHMN-like construction of self-dual string solitons using loop space. In particular, we present two extensions: The first one starts from solutions to the Basu-Harvey equation for the ABJM model, the second one starts from solutions to a corresponding BPS equation in an $\CN=2$ supersymmetric deformation of the BLG model. Both constructions yield solutions to the abelian and the nonabelian self-dual string equation transgressed to loop space. These equations might provide an effective description of M2-branes suspended between M5-branes.
\end{quote}
\end{titlepage}

\section{Introduction and results}

In recent years, problems related to finding an effective description of the M2- and M5-branes of M-theory received growing attention. In particular, Bagger-Lambert and independently Gustavsson (BLG) developed an $\CN=8$ supersymmetric Chern-Simons matter theory \cite{Bagger:2007jr,Gustavsson:2007vu}, which is a good candidate for an effective description of stacks of two M2-branes \cite{Mukhi:2008ux}. Soon after, Aharony, Bergman, Jafferis and Maldacena (ABJM) proposed a generalization of this model that is conjectured to provide an effective description of stacks of arbitrarily many M2-branes \cite{Aharony:2008ug}. In favor of this conjecture speak many results, in particular the reproduction of the peculiar $N^{3/2}$ scaling of degrees of freedom with the number $N$ of M2-branes \cite{Drukker:2010nc}.

The corresponding effective description of stacks of M5-branes, however, is much less clear. It is therefore interesting to look at a configuration of M-branes, which exhibits a duality between the M2-brane and the M5-brane theories. Recall that in type IIB superstring theory, there exists such a duality for a configuration of stacks of D1-branes ending on D3-branes. From the point of view of the D1-branes, this configuration is effectively described by the Nahm equation. The description from the perspective of the D3-branes is given by the Bogomolny monopole equation. Both are linked by the so-called Nahm transform \cite{Corrigan:1983sv,Braam:1988qk}, which maps  solutions to the Nahm equation to solutions to the Bogomolny monopole equation and vice versa. The construction of monopole solutions from solutions to the Nahm equation is also known as the Atiyah-Drinfeld-Hitchin-Manin-Nahm (ADHMN) construction \cite{Nahm:1979yw,Nahm:1981nb,Nahm:1982jb,Hitchin:1983ay}.

Lifting this D-brane configuration to M-theory, one arrives at a stack of M2-branes ending on a stack of M5-branes. The lift of the Nahm equation yields the Basu-Harvey equation \cite{Basu:2004ed}, while the lift of the Bogomolny monopole equation for gauge group $\sU(1)$ yields the self-dual string equation \cite{Howe:1997ue}. One would therefore expect an ADHMN-like construction linking solutions to the Basu-Harvey equation to self-dual string solitons. For a stack of one or two M2-branes ending on a single M5-brane, this construction was indeed found in\footnote{A different such construction using a six-dimensional auxiliary space obtained from loop space was developed in \cite{Gustavsson:2008dy}.} \cite{Saemann:2010cp}. 

Interestingly, the lift of the various components in the ADHMN construction very naturally motivates a transition to loop space\footnote{Earlier approaches to the description of M-branes on loop space are found e.g.\ in \cite{Kawamoto:2000zt,Bergshoeff:2000jn,Gustavsson:2005aq}.}, in which the self-dual string equation takes the form of a gauge theory equation. It first appears inconvenient to work with an infinite-dimensional base space, but this description has also several advantages. In particular, the self-dual string equation in its original form involves a self-dual three-form and describes only the abelian situation of a single M5-brane. On loop space, however, the corresponding gauge theory equation can be trivially rendered nonabelian and the resulting equation was conjectured in \cite{Saemann:2010cp} to describe M2-branes ending on multiple M5-branes. Further evidence for this was obtained in \cite{Papageorgakis:2011xg}: Here, a set of supersymmetric equations for a 3-Lie algebra (2,0) tensor multiplet \cite{Lambert:2010wm}, which might capture some aspects of M5-brane dynamics, was shown to have a natural interpretation on loop space. The resulting BPS equation was found to be precisely the nonabelian extension of the self-dual string equation on loop space. Moreover, the construction of \cite{Saemann:2010cp} could be straightforwardly extended to the nonabelian case.

The ADHMN-like constructions of \cite{Saemann:2010cp} and \cite{Papageorgakis:2011xg} may be conjectured to capture stacks of $k\leq 2$ M2-branes ending on arbitrarily many M5-branes. The limitation to $k\leq 2$ arises, because the constructions start from the Basu-Harvey equation based on 3-Lie algebras. In this paper, we discuss the extension to arbitrary $k$. Correspondingly, we have to switch to the BPS equation for the ABJM model, that is to a Basu-Harvey equation based on hermitian 3-algebras \cite{Bagger:2008se}. We also consider the BPS equation of a $\CN=2$ supersymmetric deformation of the BLG model based on real 3-algebras \cite{Cherkis:2008qr}. In both cases, we demonstrate how solutions to the respective Basu-Harvey equations can be used to construct solutions to the nonabelian self-dual string equation on loop space. We present various explicit examples of such solutions, and we point out their relations to corresponding solutions to the Bogomolny monopole equation in the D-brane picture. 

We also extend the constructions of \cite{Saemann:2010cp,Papageorgakis:2011xg} in another way: These constructions were formulated on the correspondence space of the transgression, which is the Cartesian product of the loop space and $S^1$. Moreover, a reduced differential operator was introduced on correspondence space to guarantee that the transgression was invertible on local abelian gerbes. Here, we work directly on loop space and we use the actual loop space exterior derivative in the construction of the gauge field strength. This leads to a slightly different self-dual string equation on loop space compared to that of \cite{Saemann:2010cp,Papageorgakis:2011xg} and it seems that in the abelian case, the loop space description of self-dual strings is richer than the direct description on space-time.

Interestingly, the fields arising in our construction take values in the gauge algebra $\au(N)_+\oplus\au(N)_-$. This gauge algebra naturally arises as the associated Lie algebra of certain hermitian 3-algebras, cf.\ appendix B. The fact that 3-algebras might underly the gauge algebra of an effective description of M5-branes has been used successfully e.g.\ in \cite{Lambert:2010wm}. The gauge algebra we find fits very well within this picture and its reinterpretation on loop space \cite{Papageorgakis:2011xg}.

There are a few open questions arising from our results. The first one concerns a quantization of $S^3$ by quantizing its loop space, cf.\ e.g.\ \cite{Saemann:2011zq}: We employ a Dirac operator containing parameterized loops in our construction. In particular, it contains the expression $\gamma^{\mu\nu}\oint \dd \tau x^\mu(\tau)\xd^\nu(\tau)$, where $x^\mu(\tau)$ with $\tau\in [0,2\pi)$ encodes a parameterized loop and $\xd(\tau)$ is the tangent vector to this loop. A homogeneity argument then suggest that the solutions to the Basu-Harvey equations used in the construction of the Dirac operator should also be dependent on the loop parameter. This would imply that these solutions form coordinates on the quantized loop space of $S^3$. These ideas should be developed in more detail, as they might also yield infinite-dimensional Euclidean 3-Lie algebras, which are not as restrictive as the finite dimensional ones.

Second, recall that by dimensionally reducing the Nahm equation and ``dimensionally oxidizing'' the Bogomolny monopole equation, one obtains\footnote{Up to certain terms in the ADHM equation.} the Nahm-dual pair appearing in the ADHM construction of instantons. It is conceivable that a similar reduction/oxidation procedure could work for the Basu-Harvey equation and the self-dual string equation on loop space, even though the M-brane interpretation is not immediately obvious.

And third, it would be interesting to ``push forward'' the interpretation of the 3-Lie algebra (2,0) tensor multiplet of \cite{Papageorgakis:2011xg} from the correspondence space to loop space. We plan to address all of these questions in future work.

This paper is structured as follows: In section 2, we begin with a review of monopoles and self-dual string solitons. We recall the ADHMN construction as well as ans\"atze for solutions to the Nahm equation and the resulting monopole configurations. We also discuss the lift of these ans\"atze to solutions to the Basu-Harvey equation. In section 3, we present the generalized Nahm transform yielding self-dual string solitons from solutions to the Basu-Harvey equation based on real 3-algebras. Section 4 then discusses the corresponding transform for the hermitian 3-algebras underlying the ABJM model. Two appendices review the definitions of generalized 3-algebras and explain our notation and a third appendix defines generalized Jacobi elliptic functions.

\section{Monopoles and self-dual strings}

\subsection{Brane interpretation}

Monopoles of charge $k$ in super Yang-Mills theory with gauge group $\sU(N)$ on $\FR^3$ can be interpreted as stacks of $k$ D1-branes ending on stacks of $N$ D3-branes in type IIB superstring theory as follows  \cite{Diaconescu:1996rk,Tsimpis:1998zh}:
\begin{equation}\label{diag:D1D3}
\begin{tabular}{rcccccccc}
& 0 & 1 & 2 & 3 & 4 & 5 & 6 & \ldots\\
D1 & $\times$ & & & & & & $\times$ \\
D3 & $\times$ & $\times$ & $\times$ & $\times$ & & &
\end{tabular}
\end{equation}
We work with Cartesian coordinates $x^0,\ldots,x^6$ on $\FR^{1,6}$ and use the identification $s=x^6$ throughout this paper. The D-brane configuration \eqref{diag:D1D3} is a BPS configuration, and the corresponding time-independent BPS equation in the effective description of the D3-branes is the {\em Bogomolny monopole equation}
\begin{equation}\label{eq:Bogomolny}
 F_{ij}=\eps_{ijk}\nabla_k\Phi~,~~~i,j,k=1,2,3~.
\end{equation}
Here, $F_{ij}$ denotes the $\au(N)$-valued curvature of the connection $\nabla_i$, and $\Phi$ is the Higgs field in the adjoint representation of $\au(N)$. The latter describes fluctuations of the D3-branes parallel to the worldvolume of the D1-branes. The time-independent BPS equation on the D1-brane, which gives rise to a dual description, is the {\em Nahm equation}
\begin{equation}\label{eq:Nahm}
 \dder{s}X^i=\tfrac{1}{2}\eps^{ijk}[X^j,X^k]~.
\end{equation}
The $X^i$ are scalar fields taking values in the adjoint of $\au(k)$. They describe the transverse fluctuations of the D1-branes parallel to the worldvolume of the D3-branes. The duality between \eqref{eq:Bogomolny} and \eqref{eq:Nahm} is a special Fourier-Mukai transform, which we discuss in some detail in section \ref{sec:ADHMNmonopoles}.

The D-brane configuration \eqref{diag:D1D3} can be lifted to M-theory by T-dualizing along the $x^5$-direction and interpreting $x^4$ as the M-theory direction. The resulting configuration is
\begin{equation}\label{diag:M2M5}
\begin{tabular}{rccccccc}
${\rm M}$ & 0 & 1 & 2 & 3 & \phantom{(}4\phantom{)} & 5 & 6 \\
M2 & $\times$ & & & & & $\times$ & $\times$ \\
M5 & $\times$ & $\times$ & $\times$ & $\times$ & $\times$ & $\times$ 
\end{tabular}
\end{equation}
This configuration is again a BPS configuration. Contrary to the case of monopoles, the corresponding BPS equation in the effective description of the M5-branes is known only for a single M5-brane, i.e.\ for $N=1$. This is the so-called {\em self-dual string equation} \cite{Howe:1997ue}
\begin{subequations}\label{eq:SelfDualString}
\begin{equation}
 H_{\mu\nu\kappa}=\eps_{\mu\nu\kappa\lambda}\dpar_\lambda\Phi~,~~~\mu,\nu,\kappa,\lambda=1,\ldots,4~,
\end{equation}
and due to the self-duality of $H$, i.e.\ $H_{\mu\nu\kappa}=\tfrac{1}{3!}\eps_{\mu\nu\kappa\rho\sigma\tau}H^{\rho\sigma\tau}$, it follows that
\begin{equation}
 H_{05\mu}=-\dpar_\mu\Phi~.
\end{equation}
\end{subequations}
As a time-independent BPS equation in the effective description of the M2-branes, Basu and Harvey \cite{Basu:2004ed} suggested the equation
\begin{equation}\label{eq:BasuHarvey}
 \dder{s}X^\mu=\tfrac{1}{3!}\eps^{\mu\nu\kappa\lambda}[X^\nu,X^\kappa,X^\lambda]~,~~~X^\mu\in\CA~,
\end{equation}
which is a natural extension of the $\sSO(3)$-symmetric Nahm equation \eqref{eq:Nahm} describing the $\sSO(3)$-symmetric configuration \eqref{diag:D1D3} to the $\sSO(4)$-symmetric situation \eqref{diag:M2M5}. Here, the $X^\mu$ are scalar fields taking values in the 3-Lie algebra\footnote{See the appendix for definitions and our conventions related to 3-algebras.} $\CA$. They describe transverse fluctuations of the M2-branes parallel to the worldvolume of the M5-branes. 

\subsection{The ADHMN construction of monopoles}\label{sec:ADHMNmonopoles}

Roughly speaking, the ADHMN construction of monopoles is a Fourier-Mukai transform over a dual pair of degenerate tori $T^4_{\rm D1}$ and $\hat{T}^4_{\rm D3}$ with radii being either infinite or zero. In the D-brane picture \eqref{diag:D1D3}, the degenerate torus $T^4_{\rm D1}=\FR^1$ corresponds to the worldvolume of the D1-branes, while its dual $\hat{T}^4_{\rm D3}=\FR^3$ is to be identified with the D3-branes' worldvolume.

To perform this transform, we start from a special solution to the Nahm equation \eqref{eq:Nahm}. Such a solution is given by a triplet of antihermitian scalar fields $X^i$ over an open interval $\CI\subsetneq \FR$ taking values in the Lie algebra $\au(k)$. Here, $\CI$ is to be identified with the spatial part of the worldvolume of the D1-branes in configuration \eqref{diag:D1D3}. The finite boundaries of $\CI$ correspond to locations of D3-branes. We demand that $X^i$ has simple poles at such finite boundary points of the interval. Moreover, the residues of the solution at these points have to form an irreducible representation of $\mathfrak{su}(2)$ of dimension $k$. 

From this solution, one constructs a Dirac operator $\nablas_{s,x}:W^{1,2}_0(\CI)\otimes\FC^2\otimes \FC^k\rightarrow W^{0,2}(\CI)\otimes\FC^2\otimes \FC^k$. Here, $W^{n,2}$ denotes the Sobolev space of functions on $\CI$, which are square integrable up to their $n$th derivative and the subscript $0$ implies that the functions vanish at finite boundaries of $\CI$, cf.\ \cite{Hitchin:1983ay}. Explicitly, the Dirac operator and its adjoint read as
\begin{equation}
 \nablas_{s,x}=-\unit\dder{s}+\sigma^i\otimes (\di X^i+x^i\unit_k)\ewith \nablabs_{s,x}:=\unit\dder{s}+\sigma^i\otimes (\di X^i+x^i\unit_k)~,
\end{equation}
where the $x^i$ are the Cartesian coordinates on $\FR^3=T^4_{\rm D3}$. Their appearance reflects the twist by the Poincar{\'e} line bundle in the Fourier-Mukai transform. The fact that the $X^i$ form a solution to the Nahm equation is equivalent to 
\begin{equation}\label{eq:RelationsNahm}
\Delta_{s,x}:=\nablabs_{s,x}\nablas_{s,x}>0 \eand [\Delta_{s,x},\sigma^i\otimes\unit_k]=0~.
\end{equation}
From the normalized zero modes $\psi^a_{s,x}\in W^{0,2}(\CI)\otimes\FC^2\otimes \FC^k$, $a=1,\ldots,N$, of $\nablabs_{s,x}$ satisfying
\begin{equation}
 \nablabs_{s,x}\psi^a_{s,x}=0~,~~~N=\dim_\FC({\rm ker}\nablabs_{s,x})\eand\delta^{ab}=\int_\CI \dd s\,\psib^a_{s,x}\psi^b_{s,x}~,
\end{equation}
one can construct the following $\au(N)$-valued gauge potential and Higgs field on $T^4_{\rm D3}$:
\begin{equation}\label{eq:BogomolnyFields}
(A_i)^{ab}:=\int_\CI \dd s\,\psib^a_{s,x}\der{x^i}\psi^b_{s,x}\eand\Phi^{ab}:=-\di\int_\CI \dd s\,\psib^a_{s,x}\,s\,\psi^b_{s,x}~.
\end{equation}
Using the relations \eqref{eq:RelationsNahm}, it is straightforward to show that these fields indeed satisfy the Bogomolny monopole equation \eqref{eq:Bogomolny}. We perform a very similar computation in the case of self-dual strings below. Two explicit examples of this construction are reviewed in section \ref{sec:ADHMNexample}.

\subsection{Examples of solutions to the Nahm equation}

For the simplest case $k=1$, the Nahm data are given by a triplet of constants $X^i\in\FR$ which describe the position of the center of mass of the monopole. In general, the components proportional to $\unit_k$ give this position, which we set to zero in this section, restricting the fields $X^i$ to $\mathfrak{su}(k)$ and fixing the center at the origin.

For $N=1$, the Nahm data live on an interval of the form $(-\infty,s_0)$ or $(s_0,\infty)$ with a simple pole at $s=s_0$. The family of spherically symmetric solutions, corresponding to $k$ coincident D1-branes ending on a single D3-brane, is given by
\begin{equation}\label{eq:sphersymm}
X^i=\frac{e^i}{s-s_0}~,
\end{equation}
where the $e^i$ form a $k$-dimensional irreducible representation of $\mathfrak{su}(2)$. 

This configuration is known as a \emph{fuzzy funnel} \cite{Constable:1999ac}: Each point of the worldvolume of the D1-brane polarizes into a fuzzy or noncommutative $S^2$ whose radius diverges at $s=s_0$. The fuzzy funnel describes a transition between D1-branes and D3-branes with a partially noncommutative worldvolume. 

To obtain more general solutions to the Nahm equations, we consider the ansatz $X^i=f_i(s) e^i$, with no sum over $i$. This ansatz was first suggested in \cite{Nahm:1981nb}, and it produces the most general solution for $k\le2$. It reduces the Nahm equations \eqref{eq:Nahm} to 
\begin{equation}\label{eq:spinningtop}
\dder{s}f_1=-f_2 f_3~,~~~
\dder{s}f_2=-f_1 f_3~,~~~ 
\dder{s}f_3=-f_1 f_2~.
\end{equation}
This system of equations is a special case of the Euler-Poinsot equations describing a spinning top in 3 dimensions. There are two constants of motion, related to the mass and energy of the spinning top: $a=f_2^2-f_1^2$ and $b=f_3^2-f_1^2$. The solutions to \eqref{eq:spinningtop} are found by substituting the constants of motion and integrating:
\begin{equation}\label{eq:generalNahm}
f_1=\frac{\sqrt b \ {\rm cn}_{{k}}(\sqrt b \ s)}{{\rm sn}_{{k}} (\sqrt b\ s)}~,~~~ 
f_2=\frac{\sqrt b \ {\rm dn}_{{k}}(\sqrt b \ s)}{{\rm sn}_{{k}} (\sqrt b\ s)}~,~~~  
f_3=\frac{\sqrt b}{{\rm sn}_{{k}} (\sqrt b\ s)}~,
\end{equation}
where $k^2=1-\frac{a}{b}$ and ${\rm cn}_k(s)$, ${\rm dn}_k(s)$ and ${\rm sn}_k(s)$ are the {\em Jacobi elliptic functions} defined in appendix \ref{app:Jacobi}. 

\begin{figure}[h]
\center
\begin{picture}(420,100)
\put(79.0,73.0){\makebox(0,0)[c]{$\dfrac{1}{s}$}}
\put(342.0,73.0){\makebox(0,0)[c]{$f_3$}}
\put(342.0,56.0){\makebox(0,0)[c]{$f_2$}}
\put(342.0,38.0){\makebox(0,0)[c]{$f_1$}}
\includegraphics[width=57mm]{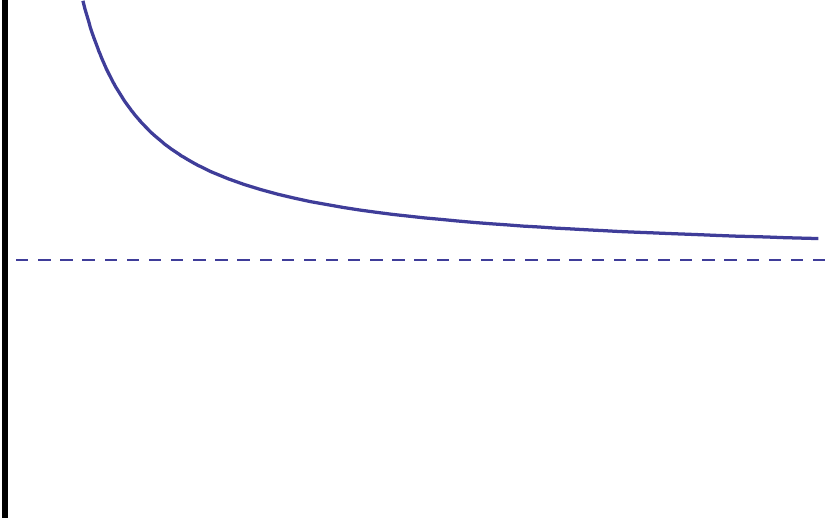}~~~~~~~~~~~~~~~~~~~~~~
\includegraphics[width=57mm]{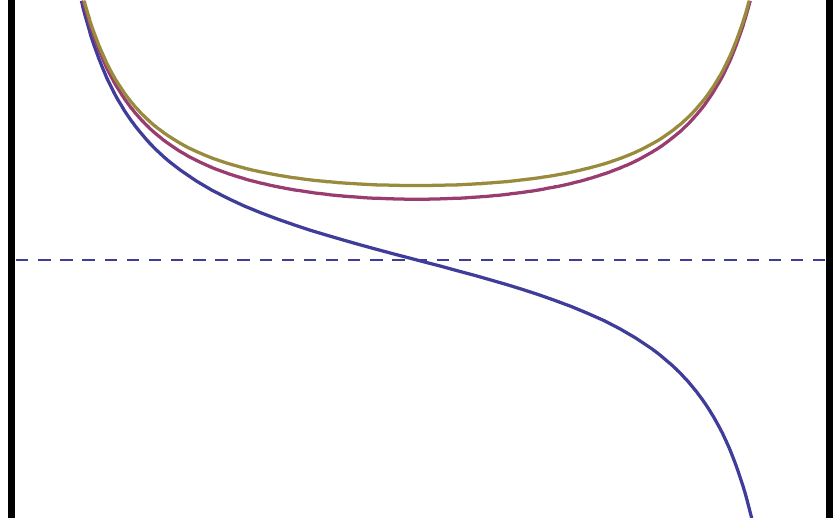}
\end{picture}
\caption{The plot on the left depicts the radial dependence $f=\frac{1}{s}$ in the spherically symmetric configuration \eqref{eq:sphersymm}. The plot on the right shows the corresponding functions $f_1(s),\ f_2(s)$ and $f_3(s)$ in \eqref{eq:generalNahm} for $a=2$, $b=3$. The vertical asymptotes give the positions of D3-branes.}
\end{figure}

The constant of integration is chosen such that one of the poles lies at $s=0$, the other lies at $s=\frac{2}{\sqrt b}{\rm sn}^{-1}_k(1)$. Note that multiplying any two functions by $-1$ gives another solution to the system, although this factor can be absorbed into the $e^i$ to give an equivalent representation of $\asu(2)$. By expanding the solutions \eqref{eq:generalNahm} around the poles, one easily shows that $X^i=\frac{e^i}{s}+$ non-singular terms.

There are two interesting special cases of solution \eqref{eq:generalNahm}. First, there is the axially symmetric case with $a=b$:
\begin{equation}
f_1=\sqrt b/{\rm tan}(\sqrt b \ s)~,~~~f_2=f_3=\sqrt b/{\rm sin}(\sqrt b \ s)~,
\end{equation}
which leads to axially symmetric non-singular monopoles for all charges $k\ge2$, cf.\  \cite{Rossi:1982fq} and references therein. Note that there are no spherically symmetric configurations for $N=2$, $k\ge2$.

Second, there is the case $a=0$, which gives $N=1$ solutions: 
\begin{equation}
f_1=f_2=\sqrt b/{\rm sinh}(\sqrt b \ s)~,~~~f_3=\sqrt b/{\rm tanh}(\sqrt b \ s)~.
\end{equation}
Here, the parameter $b$ corresponds to the separation of the monopoles. Note that the horizontal asymptotes are 0 except for $f_3$, which goes to $\sqrt b$.  Upon taking the limit $b\rightarrow 0$ we recover the spherically symmetric solution \eqref{eq:sphersymm}.

The appearance of elliptic functions is related to the fact that the Nahm equation can be formulated in terms of a Lax pair. This implies that the Nahm equation is linear on the Jacobian variety of its spectral curve \cite{Hitchin:1983ay,Adler1980318}. For the case $k=2$, the spectral curve is a torus, whose doubly-periodic complex coordinate can be identified with the complexification of the variable $s$. The Jacobi elliptic functions form a doubly-periodic basis for functions with maximally simple poles on this torus.

\subsection{Examples of monopole solutions}\label{sec:ADHMNexample}

Consider first the Nahm data \eqref{eq:sphersymm}, corresponding to a stack of $k$ coincident D1-branes ending on $N=1$ D3-branes, which we take to be located at $x^6=s=0$. The spatial part of the worldvolume of the D1-branes is thus $\CI=\FR^{>0}$. The normalized zero mode of $\nablabs_{s,x}$ at the point $\vec x=(0,0,R)^T$ is given by 
\begin{equation}\label{eq:zmMN1}
\psi=\frac{2^{\frac{k}{2}}}{\sqrt{(k-1)!}}R^{\frac{k}{2}} \de^{-s R}s^\frac{k-1}{2} (1,0,\ldots,0)^T~, 
\end{equation}
which yields the Higgs field 
\begin{equation}
\Phi=-\frac{\di k}{2R}~.
\end{equation}
For arbitrary $\vec x$, the computation of the zero modes is more difficult. Note that the Higgs field of the charge $k$ monopole is $k$ times that of a charge $1$ monopole.

Another nice example is the case of $N=2$ and $k=1$, which gives a non-singular $\sSU(2)$ monopole. The Nahm data are constants, taken to be $0$ and the interval is taken to be $(-s_0,s_0)$. The normalized zero modes, in matrix notation, are given by
\begin{equation}
\psi=\sqrt{\frac{R}{{\rm sinh}(2s_0R)}}({\rm cosh}(Rs)\unit +{\rm sinh}(Rs)\frac{x^i\sigma^i}{R})~,
\end{equation}
which yields the non-singular Higgs field 
\begin{equation}
\Phi=\frac{\di x^i\sigma^i}{R^2}(s_0R\ {\rm coth}(s_0R)-1)~.
\end{equation}

\subsection{Examples of solutions to the Basu-Harvey equation}

The Basu-Harvey equation \eqref{eq:BasuHarvey} also has a unique $\sSO(4)$ invariant, $N=1$ solution given by $X^\mu= \frac{e^\mu}{\sqrt{2(s-s_0)}}$, where the $e^\mu$ are generators of the 3-Lie algebra $A_4$: $[e^\mu,e^\nu,e^\kappa]=\epsilon^{\mu\nu\kappa\lambda}e^\lambda$. This corresponds to a stack of two coincident M2-branes ending on a single M5-brane and, analogously to the D1-D3-brane configuration, a fuzzy funnel (of one higher dimension) is believed to occur \cite{Basu:2004ed}.

Similarly to the previous ansatz for the Nahm equation, the ansatz $X^\mu=f_\mu (s)e^\mu$ (no sum over $\mu$ implied) reduces the Basu-Harvey equation \eqref{eq:BasuHarvey} to
\begin{equation}\label{eq:3AspinningTop}
\begin{aligned}
\dder{s}f_1&=-f_2 f_3 f_4~,~~~
\dder{s}f_3&=-f_1 f_2f_4~,~~~
\dder{s}f_2&=-f_1 f_3f_4~,~~~
\dder{s}f_4&=-f_1 f_2 f_3~.
\end{aligned}
\end{equation}
The constants of motion for this system are\footnote{As usual in Nambu mechanics \cite{Takhtajan:1993vr}, where the 
Poisson bracket is replaced by a Nambu bracket with 3 arguments, one has an extra Hamiltonian and hence an extra constant of motion.} $a=f_2^2-f_1^2,\ b=f_3^2-f_1^2$ and $c=f_4^2-f_1^2$. The solutions to \eqref{eq:3AspinningTop} were first found in \cite{Nogradi:2005yk}. They are given by \emph{generalized Jacobi elliptic functions}, which are hyperelliptic but can be viewed as single-valued meromorphic functions on a Riemann surface of genus two \cite{Pawellek:2009er}. Using \eqref{genjac}, the solutions can be expressed in terms of Jacobi elliptic functions  
\begin{equation}\label{eq:generalBH}
\begin{aligned}
f_1&=-\frac{\sqrt{a}\ {\rm sn}_{\kappa}( ps)}{\sqrt{1-\frac{a}{c}-{\rm sn}_{\kappa}^2(ps)}}~,~~~
&f_3&=\frac{\sqrt{b(1-\frac{a}{c})}\ {\rm dn}_{\kappa}( ps)}{\sqrt{1-\frac{a}{c}-{\rm sn}_{\kappa}^2(ps)}}~,\\
f_2&=\frac{\sqrt{a(1-\frac{a}{c})}}{\sqrt{1-\frac{a}{c}-{{\rm sn}_{\kappa}^2(ps)}}}~,~~~
&f_4&=\frac{\sqrt{c-a}\ {\rm cn}_{\kappa}( ps)}{\sqrt{1-\frac{a}{c}-{\rm sn}_{\kappa}^2(ps)}}~,
\end{aligned}
\end{equation}
where $p^2=b(c-a)$ and $\kappa^2=\frac{c(b-a)}{b(c-a)}$. This solution exhibits singular behavior at $s_0=\pm \frac{1}{p}{\rm sn}^{-1}_{\kappa'}(\sqrt{1-\frac{a}{c}})$. Expanding around these points by using the identities \eqref{C1}, we see that $X^\mu\sim \frac{e^\mu}{\sqrt{2(s-s_0)}}+$ non-singular terms.
\begin{figure}[h]
\center
\begin{picture}(420,100)
\put(79.0,82.0){\makebox(0,0)[c]{$\frac{1}{\sqrt{2s}}$}}
\put(342.0,90.0){\makebox(0,0)[c]{$f_4$}}
\put(399.0,85.0){\makebox(0,0)[c]{$f_3$}}
\put(342.0,62.0){\makebox(0,0)[c]{$f_2$}}
\put(342.0,35.0){\makebox(0,0)[c]{$f_1$}}
\includegraphics[width=57mm]{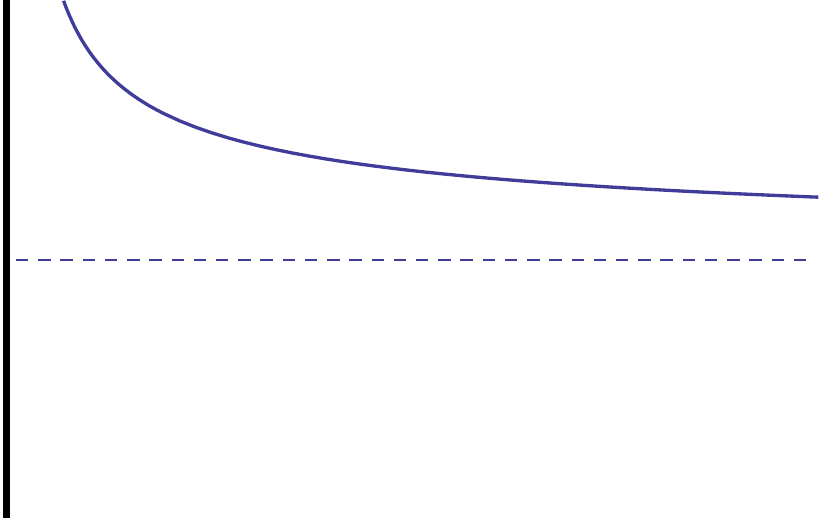}~~~~~~~~~~~~~~~~~~~~~~
\includegraphics[width=57mm]{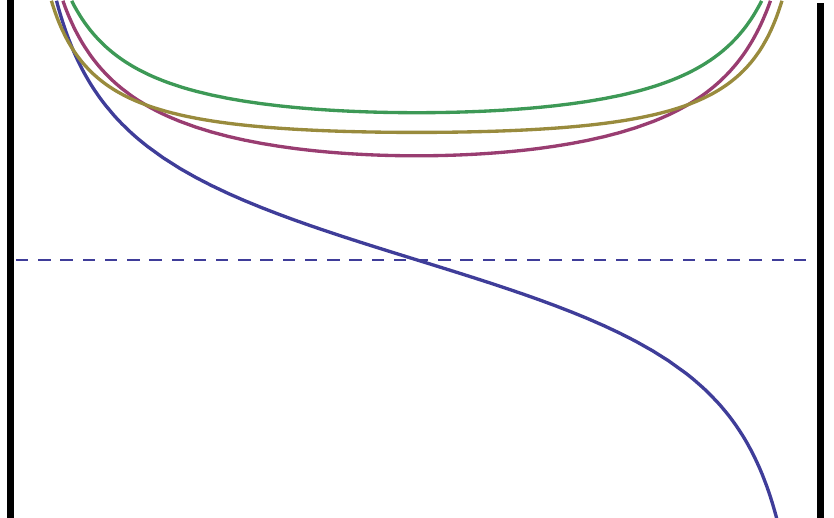}
\end{picture}
\caption{The plot on the left depicts the radial dependence $1/\sqrt{2(s-s_0)}$ of the $N=1$ solution. The plot on the right shows the corresponding functions $f_1(s),\ f_2(s),\ f_3(s)$ and $f_4(s)$ of the solution \eqref{eq:generalBH} for $a=2$, $b=3$, $c=4$. The vertical asymptotes give the positions of the M5-branes.}
\end{figure}
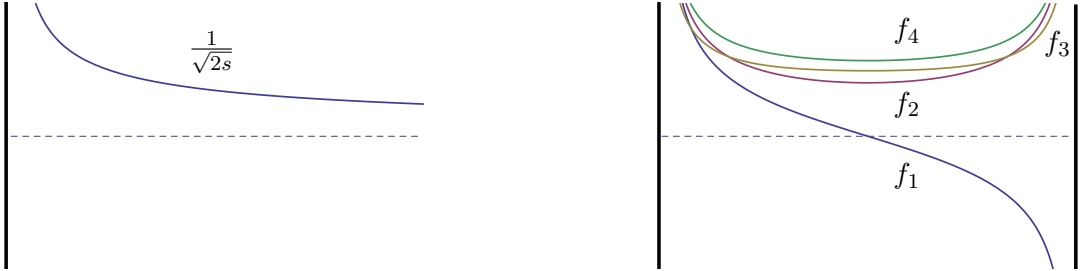

We can again take two interesting limits of the solution \eqref{eq:generalBH}. First, there is the axially symmetric case for $a=b=c$:
\begin{equation}
f_1=-b\ s~\sqrt{\frac{b}{1-b^2s^2}}~,~~~f_2=f_3=f_4=\sqrt{\frac{b}{1-b^2s^2}}~.
\end{equation}
Second, the limit $a\rightarrow0$ takes the period to infinity, giving $N=1$ solutions:
\begin{equation}
\begin{aligned}
f_1=f_2=\frac{ p }{\sqrt{{\rm sinh}(ps)(2p{\rm cosh}(ps)+(b+c){\rm sinh}(ps))}}~,\hspace{2.5cm}&\\
f_3=\frac{ p + b\ {\rm tanh}(ps)}{\sqrt{{\rm tanh}(ps)(2p+(b+c){\rm tanh}(ps))}}~,~
f_4=\frac{ p + c\ {\rm tanh}(ps)}{\sqrt{{\rm tanh}(ps)(2p+(b+c){\rm tanh}(ps))}}~,&
\end{aligned}
\end{equation}
where $p^2=bc$. Taking $b\rightarrow0$ then gives
\begin{equation}
f_1=f_2=f_3=\frac{1}{\sqrt{s(2+cs)}}~,~~~f_4=\frac{1+cs}{\sqrt{s(2+cs)}}~.
\end{equation}
The horizontal asymptotes are now 0 except for $f_4$, which goes to $\sqrt c$. Taking the parameter $c\rightarrow0$ gives the $\sSO(4)$ symmetric case $X^\mu=\frac{e^\mu}{\sqrt{2s}}$ as expected.

\subsection{Self-dual strings on loop space}

It is not clear how to perform an ADHMN-like construction for self-dual string solitons directly. However, one can reformulate the self-dual string equation \eqref{eq:SelfDualString} on loop space, for which such a construction has been found in \cite{Saemann:2010cp}.

Just as a Dirac monopole is described by the first Chern class $F\in H^2(M,\RZ)$ of a principal $\sU(1)$-bundle over the manifold $M=\FR^3$ or rather\footnote{Dirac monopole solutions on $\FR^3$ are singular at the position of the monopoles, and one should therefore consider the principal $\sU(1)$-bundle on a sphere with the monopole at its center.} $M=S^2$, a self-dual string can be described by the Dixmier-Douady class $H\in H^3(M,\RZ)$ of an abelian $\sU(1)$-gerbe over the manifold $M=S^3$. Working with three-form field strengths is rather inconvenient, but there is a trick which allows us to map the Dixmier-Douady class to a first Chern class. This map is called a {\em transgression} \cite{0817647309} and it is defined as follows: Consider $k$ vector fields $v_1,\ldots, v_k$ on the loop space $\CL M$ of $M$. In components, we have
\begin{equation}
v_i=\oint \dd \tau\, v_i^\mu(\tau)\delder{x^\mu(\tau)}~. 
\end{equation}
Any $k+1$-form $\omega\in \Omega^{k+1}(M)$ on $M$ is mapped to a $k$-form $\CT\omega\in\Omega^k(\CL M)$ via
\begin{equation}
 (\CT\omega)_x(v_1(x),\ldots,v_k(x)):=\oint_{S^1}\dd\tau\,\omega(v_1(\tau),\ldots,v_k(\tau),\dot{x}(\tau))~.
\end{equation}
Here, $x\in \CL M$ denotes a loop and $\xd(\tau)$ is the tangent vector to the loop $x$ at $\tau$. By going to loop space, we thus gain a natural vector, which we can use to fill up one slot of a differential form. Note that the price we have to pay for using the transgression map $\CT$ is that we are now working with an infinite-dimensional base space. One can readily check that $\CT$ is a chain map. This implies that given a three-form field strength $H=\dd B$ of a two-form potential $B$ on $M$, $F=\CT H$ is indeed the field strength for the gauge potential $A=\CT B$ on $\CL M$.

The transgression of the self-dual string equation \eqref{eq:SelfDualString} is given in \cite{Saemann:2010cp} by
\begin{equation}\label{eq:LoopSpaceSDS}
 F_{\mu\nu}(\tau)=\eps_{\mu\nu\kappa\lambda}\xd^\kappa(\tau) \dpar_\lambda\Phi(x(\tau))~,
\end{equation}
where $F$ is a $\au(1)$-valued curvature of some gauge potential, $\Phi$ is a Higgs field and the loop space derivative is
\begin{equation}\label{eq:redloopderivative}
 \dpar_\mu:=\oint_{S^1}\dd \sigma\,\frac{\delta}{\delta x^\mu(\sigma)}~.
\end{equation}

Note that, since the loop parameter $\tau$ appears explicitly in \eqref{eq:LoopSpaceSDS}, this equation does not live on loop space but on the {\it correspondence space} $\CL S^3\times S^1$. In particular, the Higgs field $\Phi(x(\tau))$ is the pullback of the Higgs field $\Phi(x)$ on $S^3$ along the evaluation map $ev:\CL S^3\times S^1\rightarrow S^3: (x(\tau),\tau_0)\mapsto x(\tau_0)$. Here, we intend to perform the construction on loop space itself. That is, we use the loop space exterior derivative 
\begin{equation}
\delta:=\oint \dd \sigma\, \delta x^{\mu}(\sigma)\wedge \delta_{(\mu\sigma)}\ewith \delta_{(\mu\sigma)}:=\delder{x^\mu(\sigma)}~,
\end{equation}
and we consider a Higgs field $\Phi$, which is a $\au(1)$-valued function on $\CL S^3$. Such a function $\Phi$ can be derived from a Higgs field $\Phi_{S^3}$ on $S^3$ by a transgression of functions, i.e.\ via pull-back to the correspondence space and subsequent integration: $\Phi=\oint_{S^1}\dd \tau\, |\xd(\tau)|\,\Phi_{S^3}(x(\tau))$. Moreover, we allow for arbitrary gauge potentials $A$ on $\CL S^3$, which are not necessarily of the form $\CT B$ for some two-form potential $B$ on $S^3$.

Note that a general field strength on loop space is of the form
\begin{equation}
\begin{aligned}
 F:=&\delta A:=\oint \dd \sigma\, \delta x^{\mu}(\sigma)\wedge\delder{x^\mu(\sigma)}\oint \dd \tau\, \delta x^{\nu}(\tau) A_{(\nu\tau)}\\
=&\oint \dd \sigma \oint \dd \tau\, F_{(\mu\sigma)(\nu\tau)}\delta x^{\mu}(\sigma)\wedge \delta x^{\nu}(\tau)~,
\end{aligned}
\end{equation}
where
\begin{equation}
 F_{(\mu\sigma)(\nu\tau)}:=\delder{x^\mu(\sigma)}A_{(\nu\tau)}-\delder{x^\nu(\tau)}A_{(\mu\sigma)}~.
\end{equation}
In equation \eqref{eq:LoopSpaceSDS}, however, only an {\em ultra-local} expression appears. That is, the field strength is of the form
\begin{equation}\label{eq:ultralocal}
 F_{(\mu\sigma)(\nu\tau)}=F_{\mu\nu}(\tau)\delta(\sigma-\tau)~.
\end{equation}

This implies, that we have to extend the self-dual string equation to get both the terms antisymmetric in $\mu\nu$ (and correspondingly symmetric in $\tau\sigma$) as well as the terms symmetric in $\mu\nu$ (and correspondingly antisymmetric in $\tau\sigma$). The extension of \eqref{eq:LoopSpaceSDS} that appears in our construction is given by
\begin{equation}\label{eq:exLoopSpaceSDS}
 F_{(\mu\sigma)(\nu\tau)}=\left(\eps_{\mu\nu\kappa\lambda}\xd^\kappa(\sigma)\delder{x^\lambda(\tau)}\Phi\right)_{(\sigma\tau)}-\Gamma_{\rm ch}\left(2\xd_{(\mu}(\sigma)\delder{x^{\nu)}(\tau)}\Phi-\delta_{\mu\nu}\xd^\kappa(\sigma)\delder{x^\kappa(\tau)}\Phi\right)_{[\sigma\tau]}~,
\end{equation}
where $(\cdot)_{(\sigma\tau)}$ and $(\cdot)_{[\sigma\tau]}$ denote symmetrization and antisymmetrization in $\sigma$ and $\tau$, respectively. The fields $F_{(\mu\sigma)(\nu\tau)}$ and $\Phi$ now take values in the abelian Lie algebra $\frg=\au(1)_+\oplus\au(1)_-$ and $\Gamma_{\rm ch}$ is a linear involution on $\frg$ with $\Gamma_{\rm ch} (\lambda_\pm)=\pm\lambda_\pm$ for $\lambda_\pm\in\au_\pm(1)$. The obvious nonabelian generalization of the self-dual string equation on loop space \eqref{eq:exLoopSpaceSDS} is:
\begin{multline}\label{eq:NALoopSpaceSDS}
 F_{(\mu\sigma)(\nu\tau)}=\big(\eps_{\mu\nu\kappa\lambda}\xd^\kappa(\sigma)\nabla_{(\lambda\tau)}\Phi\big)_{(\sigma\tau)}\\-\Gamma_{\rm ch}\big(\xd_{\mu}(\sigma)\nabla_{(\nu\tau)}\Phi+\xd_{\nu}(\sigma)\nabla_{(\mu\tau)}\Phi-\delta_{\mu\nu}\xd^\kappa(\sigma)\nabla_{(\kappa\tau)}\Phi\big)_{[\sigma\tau]}~.
\end{multline}
where $\nabla_{(\mu\sigma)}=\delta_{(\mu\sigma)}+A_{(\mu\sigma)}$, the fields take values in $\frg=\au(N)_+\oplus\au(N)_-$ and $\Gamma_{\rm ch}(\lambda_\pm):=\pm\lambda_\pm$ for $\lambda_\pm\in\au_\pm(N)$.

The physical interpretation of this equation is yet unclear: Assuming that a self-dual string is fully described in terms of equation \eqref{eq:SelfDualString}, the components of \eqref{eq:exLoopSpaceSDS} antisymmetric in $\sigma$ and $\tau$ are superfluous, as they cannot be obtained from \eqref{eq:SelfDualString} by a transgression map. Indeed, without the terms antisymmetric in $\sigma\tau$, we have the unextended nonabelian self-dual string equation on loop space \cite{Gustavsson:2008dy,Saemann:2010cp,Papageorgakis:2011xg}.  In \cite{Papageorgakis:2011xg}, this reduced form of equation \eqref{eq:exLoopSpaceSDS} was shown to be the BPS equation to a loop space interpretation of the 3-Lie algebra (2,0) tensor multiplet equations of \cite{Lambert:2010wm}. Note however, that the transgression of \eqref{eq:SelfDualString} is contained in \eqref{eq:exLoopSpaceSDS}. In the abelian case, where the equation is linear, we can therefore project from solutions of \eqref{eq:exLoopSpaceSDS} onto solutions of the transgression of \eqref{eq:SelfDualString}. Moreover, equation \eqref{eq:exLoopSpaceSDS} appears naturally in the Nahm-like construction on loop space, which we develop in the following section. This also motivates the generalization to gauge algebra $\frg$: The Nahm-like construction starts from 3-algebras that often come with an associated Lie algebra of the form $\au(N)_+\oplus\au(N)_-$, which induces a similar splitting onto the constructed fields.

Note that strictly speaking, one should replace $\xd^\rho$ by $R\xd^\rho/|\xd|$, $R\in\FR^{>0}$, in \eqref{eq:NALoopSpaceSDS} and in all of our other equations to arrive at equations invariant under reparameterizations of the loops. To simplify notation, we refrain from doing this but fix the parameterization of all loops by demanding that $\xd^\mu(\tau)\xd^\mu(\tau)=R^2$. 

If the fields take values in $\au(N)_+\oplus\au(N)_-$ with $N>1$, then the Higgs field does not have to diverge and we can extend our considerations from the loop space $\CL S^3$ to the loop space of $\FR^4$.

In the rest of the paper, we are concerned with constructing various solutions to equations \eqref{eq:NALoopSpaceSDS} by using an ADHMN-like construction.

\section{Self-dual strings from real 3-algebras}

The original construction of self-dual strings developed in \cite{Saemann:2010cp} made use of 3-Lie algebras and the restricted loop space derivative $\dpar_\mu$. Here, we present the extension involving real 3-algebras and the loop space exterior derivative $\delta$. Recall that all 3-Lie algebras are special cases of real 3-algebras, cf.\ appendix \ref{app:real3algebras}. 

\subsection{The Basu-Harvey equation for real 3-algebras}

The Basu-Harvey equation \eqref{eq:BasuHarvey} is a BPS equation in the BLG model in which the matter fields take values in a 3-Lie algebra and the gauge potential lives in the associated Lie algebra. The problem with using 3-Lie algebras is that they are highly restricted: the only finite-dimensional 3-Lie algebras with positive definite metric are $A_4$ and direct sums thereof. In \cite{Cherkis:2008qr,Cherkis:2008ha}, it was therefore suggested to consider the BLG model with matter fields valued in a real 3-algebra, which preserves at least $\CN=2$ supersymmetries. Another, more interesting generalization of 3-Lie algebras is given by the hermitian 3-algebras, to which we come in section \ref{sec:h3}.

From the supersymmetry transformations given in \cite{Cherkis:2008ha}, it is straightforward to derive the BPS equation corresponding to the Basu-Harvey equation for real 3-algebras. With appropriate normalization, the result is just the ordinary Basu-Harvey equation with the fields $X^\mu$ taking values in a real 3-algebra:
\begin{equation}\label{eq:r3BasuHarvey}
 \dder{s}X^\mu=\tfrac{1}{3!}\eps^{\mu\nu\kappa\lambda}[X^\nu,X^\kappa,X^\lambda]~,~~~X^\mu\in\CA~.
\end{equation}
A class of examples of real 3-algebras is given in the appendix. In particular, the 3-Lie algebra $A_4$ is a sub 3-algebra of the real 3-algebra $\CCC_4$. 

\subsection{The construction}

Analogously to the case of the ADHMN construction, we start from a Dirac operator built from a solution to the Basu-Harvey equation \eqref{eq:r3BasuHarvey}. The solution consists of a quadruplet of real scalar fields over the interval $\CI$ which take values in a metric real 3-algebra $\CA$. Contrary to the case of monopoles, where the solution to the Nahm equation had to have a simple pole at finite boundaries $s_0$ of $\CI$, we demand here that
\begin{equation}
 X^\mu(s)\sim\frac{e^\mu}{\sqrt{2(s-s_0)}}+\mbox{regular terms}~.
\end{equation}
The Dirac operator is a map $\nablas_{s,x}:W^{1,2}_0(\CI)\otimes\FC^4\otimes \CA\rightarrow W^{0,2}(\CI)\otimes\FC^4\otimes \CA$ and explicitly, we have
\begin{equation}\label{eq:r3DiracOperators}
\begin{aligned}
 \nablas_{s,x}&=-\gamma_5\dder{s}+\tfrac{1}{2}\gamma^{\mu\nu}\left(D(X^\mu,X^\nu)+\di \oint \dd \tau\, x^\mu(\tau)\xd^\nu(\tau)\right)~,\\
 \nablabs_{s,x}&=+\gamma_5\dder{s}+\tfrac{1}{2}\gamma^{\mu\nu}\left(D(X^\mu,X^\nu)+\di \oint \dd \tau\, x^\mu(\tau)\xd^\nu(\tau)\right)~.
\end{aligned}
\end{equation}
A detailed motivation for the form of this Dirac operator is found in \cite{Saemann:2010cp}. The expressions $x^{\mu\nu}:=\oint \dd \tau\, x^\mu(\tau)\xd^\nu(\tau)$ are also known as the {\em area coordinates} or integrated {\em Pl\"ucker coordinates} of the loop $x$.\footnote{Due to $x^{\mu\nu}=\oint \dd \tau\, x^\mu(\tau)\xd^\nu(\tau)=\oint_C x^\mu\dd x^\nu=\int_V\dd x^\mu\wedge \dd x^\nu$, where $\dpar V=C$, the functions $x^{\mu\nu}$ on $\CL\FR^4$ measure the ``shadow'' of the loop projected onto the coordinate plane $\mu,\nu$.} Since the $X^\mu$ satisfy the Basu-Harvey equation, the Laplace operator $\Delta_{s,x}:=\nablabs_{s,x}\nablas_{s,x}$ is positive and commutes with the generators of $\sSpin(4)$:\footnote{Recall that in the Nahm construction, positivity of the Laplace operator was equivalent to the Dirac operator being constructed from solutions to the Nahm equation. Here, the Laplace operator is positive, if the Dirac operator is constructed from solutions to the Basu-Harvey equation. The inverse statement is only true if the map $D:\CA\wedge \CA\rightarrow \frg_\CA$ is nondegenerate, which is not the case in general.}
\begin{equation}\label{eq:PropDeltaM}
 \Delta_{s,x}>0~,~~~[\Delta_{s,x},\gamma^{\mu\nu}]=0~.
\end{equation}
Note that these properties are preserved, if we shift the Dirac operator by
\begin{equation}\label{eq:shiftDiracReal}
 \nablas_{s,x}\rightarrow \nablas_{s,x}+\gamma^{\mu\nu}\oint \dd \tau\, X^{\mu}_{0}(\tau)\xd^\nu(\tau)\ewith
 \nablabs_{s,x}\rightarrow \nablabs_{s,x}+\gamma^{\mu\nu}\oint \dd \tau\, X^{\mu}_{0}(\tau)\xd^\nu(\tau)~,
\end{equation}
where the field $X^{\mu}_0(\tau)=\di x^\mu_0(\tau)\id_\CA$ with $x^\mu_0\in \CL\FR^4$ allows for a center of mass motion of the self-dual string. For the moment, let us put $x^\mu_0=0$ to simplify the discussion.

We start from the normalized zero modes $\psi^a_{s,x}$ satisfying
\begin{equation}\label{eq:normalizePsiM}
 \nablabs_{s,x}\psi^a_{s,x}=0\eand\delta^{ab}=\int_\CI\dd s\, (\psib^a_{s,x},\psi^b_{s,x})~,
\end{equation}
where $(\dotsp,\dotsp)$ denotes the inner product on $\FC^4\otimes\CA$. We sort the zero modes according to their chirality: We have $N$ zero modes $\psi^a_{s,x}$, $a=1,\ldots,N$, with $\gamma_5\psi^a_{s,x}=\psi^a_{s,x}$ and $N$ zero modes $\psi^a_{s,x}$, $a=N+1,\ldots,2N$, with $\gamma_5\psi^a_{s,x}=-\psi^a_{s,x}$. This is possible because of the block-diagonal structure of the Dirac operator \eqref{eq:r3DiracOperators}.

Analogously to the ADHMN construction, we introduce the following fields:
\begin{equation}\label{eq:MFieldDefinitions}
 A_{(\mu\tau)}^{ab}=\int \dd s\, \left(\psib^a_{s,x}, \delder{x^\mu(\tau)} \psi^b_{s,x}\right)\eand\Phi^{ab}=\di\int \dd s\, \left(\psib^a_{s,x},\,s\,\psi^b_{s,x}\right)~.
\end{equation}
These fields are manifestly anti-hermitian and our sorting of zero modes implies that the fields take values in the gauge algeba $\au(N)_+\oplus\au(N)_-$. Note that the components in $\au(N)_\pm$ depend only on the (anti)-self-dual parts of $D(X^\mu,X^\nu)\pm\tfrac{1}{2}\eps^{\mu\nu\kappa\lambda}D(X^\kappa,X^\lambda)$. 

Let us quickly verify that these fields indeed satisfy the self-dual string equation on loop space \eqref{eq:NALoopSpaceSDS}. For this, we introduce the Green's function $G_{x}(s,t)$ which we can define via
\begin{equation}
\Delta_{s,x} G_{x}(s,t)=-\delta(s-t)
\end{equation}
due to \eqref{eq:PropDeltaM}. We then have the following completeness relation:
\begin{equation}\label{eq:completeness}
 \delta(s-t)=\psi^a_{s,x}\big(\psib^a_{t,x},\dotsp\big)-\nablas_{s,x} G_x(\tau)(s,t)\nablabs_{t,x}~.
\end{equation}
This relation, together with equation \eqref{eq:PropDeltaM} and the identities\footnote{Here and in the following, the sign $\stackrel{[\cdot]}{=}$ means that equality holds after antisymmetrizing the multi-indices $\mu\sigma$ and $\nu\tau$, i.e.\ after the expressions are antisymmetrized in $\mu\nu$ and symmetrized in $\sigma\tau$ or vice versa. We include weight factors in all symmetrizations and antisymmetrizations.}
\begin{equation*}
\begin{aligned}
\gamma^{\mu\kappa}\gamma^{\nu\lambda}\xd^\kappa(\sigma)\xd^\lambda(\tau)\stackrel{[\cdot]}{=}2\gamma^{\mu\lambda}\xd^\nu(\sigma)\xd^\lambda(\tau)-&\delta^{\mu\nu}\gamma^{\kappa\lambda}\xd^\kappa(\sigma)\xd^\lambda(\tau)+\eps_{\mu\nu\kappa\lambda}\gamma^{\kappa\rho}\gamma_5\xd^\lambda(\sigma)\xd^\rho(\tau)~,\\
 \int\dd s\left(\delder{x^\mu(\tau)}\psib^a_{s,x},\psi^b_{s,x}\right)+&\int\dd s\left(\psib^a_{s,x},\delder{x^\mu(\tau)}\psi^b_{s,x}\right)=0~,\\
\left(\delder{x^\mu(\tau)}\nablabs_{s,x}\right) \psi^a_{s,x}+&\nablabs_{s,x}\delder{x^\mu(\tau)}\psi^a_{s,x}=0~,~~~\\
\delder{x^\mu(\tau)}\nablas_{s,x}=\delder{x^\mu(\tau)}\tfrac{\di}{2}\gamma^{\kappa\lambda}&\oint \dd \sigma\,x^\kappa(\sigma)\xd^\lambda(\sigma)=\di\gamma^{\mu\lambda}\xd^\lambda(\tau)~.
\end{aligned}
\end{equation*}
allows us to compute
\begin{equation*}
 \begin{aligned}
  F_{(\mu\sigma)(\nu\tau)}^{ab}&\stackrel{[\cdot]}{=}2\int_\CI \dd s\,\big(\delta_{(\mu\sigma)}\psib^a_{s,x},\delta_{(\nu\tau)}\psi^b_{s,x}\big)+2\int_\CI \dd s\int_\CI \dd t\,\big(\psib^a_{s,x},\delta_{(\mu\sigma)}\psi^c_{s,x}\big)\big(\psib^c_{t,x},\delta_{(\nu\tau)}\psi^b_{t,x}\big)\\
&\stackrel{[\cdot]}{=}-2\int_\CI \dd s\int_\CI \dd t\,\Big(\delta_{(\mu\sigma)}\psib^a_{s,x}\,,\,\left(\nablas_{s,x}G_x(s,t)\nablabs_{t,x}\right)\delta_{(\nu\tau)}\psi^b_{t,x}\Big)\\
&\stackrel{[\cdot]}{=}2\int_\CI \dd s\int_\CI \dd t\,\Big(\psib_{s,x}^a,\left(\gamma^{\mu\kappa}\xd^\kappa(\sigma) G_x(s,t)\gamma^{\nu\lambda}\xd^\lambda(\tau)\right)\psi^b_{t,x}\Big)\\
&\stackrel{[\cdot]}{=}2\eps_{\mu\nu\kappa\lambda}\int_\CI \dd s\int_\CI \dd t\,\Big(\psib_{s,x}^a,\,G_x(s,t)\gamma^{\kappa\rho}\gamma_5\xd^\lambda(\sigma)\xd^\rho(\tau)\psi^b_{t,x}\Big)\\
&\hspace{0.5cm}+\int_\CI \dd s\int_\CI \dd t\,\Big(\psib_{s,x}^a,\,G_x(s,t)\left(4\gamma^{\mu\lambda}\xd^\nu(\sigma)\xd^\lambda(\tau)-2\delta^{\mu\nu}\gamma^{\kappa\lambda}\xd^\kappa(\sigma)\xd^\lambda(\tau)\right)\psi^b_{t,x}\Big)~.\\
 \end{aligned}
\end{equation*}
It is here that we use the fact that, since the Dirac operator is block diagonal, $\psi^b_{t,x}$ can be arranged into N left and N right-handed zero-modes. Therefore $\psi^b_{t,x}=\gamma_5\Gamma_{\rm ch}{}^b{}_c\psi^c_{t,x}$ where $\Gamma_{\rm ch}$ denotes\footnote{By a slight abuse of notation, we denote the linear involution $\Gamma_{\rm ch}$ on the gauge algebra and the matrix $\diag(\unit_N,-\unit_N)$ leading to it by the same symbol.} the diagonal matrix $\diag(\unit_N,-\unit_N)$.
\begin{equation*}
 \begin{aligned}
F_{(\mu\sigma)(\nu\tau)}^{ab}&\stackrel{[\cdot]}{=}\di\eps_{\mu\nu\kappa\lambda}\xd^\kappa(\sigma)\int_\CI \dd s\,\Big((\nabla_{(\lambda\tau)}\psib_{s,x})^a,\,s\,\psi^b_{s,x}\Big)+\Big(\psib^a_{s,x},\,s\,(\nabla_{(\lambda\tau)}\psi_{s,x})^b\Big)\\
&\hspace{0.5cm}-2\di\xd_{\mu}(\sigma)\int_\CI \dd s\,\Big((\nabla_{(\nu\tau)}\psib_{s,x})^a,\,s\,\psi^c_{s,x}\Big)\Gamma_{\rm ch}{}^b{}_c+\Big(\psib^a_{s,x},\,s\,(\nabla_{(\nu\tau)}\psi^c_{s,x}\Big)\Gamma_{\rm ch}{}^b{}_c\\
&\hspace{0.5cm}-2\di\xd_{\nu}(\sigma)\int_\CI \dd s\,\Big((\nabla_{(\mu\tau)}\psib_{s,x})^a,\,s\,\psi^c_{s,x}\Big)\Gamma_{\rm ch}{}^b{}_c+\Big(\psib^a_{s,x},\,s\,(\nabla_{(\mu\tau)}\psi^c_{s,x}\Big)\Gamma_{\rm ch}{}^b{}_c\\
&\hspace{0.5cm}+\di\delta_{\mu\nu}\xd^\kappa(\sigma)\int_\CI \dd s\,\Big((\nabla_{(\kappa\tau)}\psib_{s,x})^a,\,s\,\psi^c_{s,x}\Big)\Gamma_{\rm ch}{}^b{}_c+\Big(\psib^a_{s,x},\,s\,(\nabla_{(\kappa\tau)}\psi^c_{s,x}\Big)\Gamma_{\rm ch}{}^b{}_c\\
&\stackrel{[\cdot]}{=}\Big(\eps_{\mu\nu\kappa\lambda}\xd^\kappa(\sigma)\nabla_{(\lambda\tau)}\Phi-\Gamma_{\rm ch}(\xd_{\mu}(\sigma)\nabla_{(\nu\tau)}\Phi+\xd_{\nu}(\sigma)\nabla_{(\mu\tau)}\Phi-\delta_{\mu\nu}\xd^\kappa(\sigma)\nabla_{(\kappa\tau)}\Phi)\Big)^{ab}.
 \end{aligned}
\end{equation*}
Thus, the fields \eqref{eq:MFieldDefinitions} indeed satisfy the self-dual string equation on loop space \eqref{eq:NALoopSpaceSDS}.

\subsection{Comments on the reduction to monopoles}\label{sec:r3reduction}

The duality between solutions to the nonabelian self-dual string equation on loop space \eqref{eq:NALoopSpaceSDS} and solutions to the Basu-Harvey equation \eqref{eq:r3BasuHarvey} can be reduced to the duality between solutions to the Bogomolny monopole equation and solutions to the Nahm equation. This reduction has been explained in detail in \cite{Saemann:2010cp} and \cite{Papageorgakis:2011xg} for 3-Lie algebras, and the transition to real 3-algebras is trivially performed. Let us therefore just summarize the key steps in the following.

As usual when going from M-theory to string theory, we have to compactify spacetime along an M-theory direction, which we choose here to be the $x^4$-direction. That is, we arrive at the loop space of $\FR^3\times S^1$ and the radius of the contained $S^1$ is identified with $R=g_{\rm YM}^2=\frac{1}{2\pi}$. We restrict ourselves to loops wrapping this circle by demanding $x^\mu(\tau)=x_0^\mu+R\delta_4^\mu\tau$ and thus $\xd^\mu=R\delta_4^\mu$. In the Dirac operator \eqref{eq:r3DiracOperators}, the generators $\gamma^{\mu\nu}$ of $\sSpin(4)$ are reduced to $\gamma^{i4}$, which generate $\sSU(2)\cong \sSpin(3)\subset \sSpin(4)$. Moreover, because the area coordinates reduce according to
\begin{equation}
 \tfrac{1}{2}\oint \dd \tau\, \gamma^{\mu\nu}x^\mu(\tau)\xd^\nu(\sigma)=\gamma^{i4}x_0^i~,
\end{equation}
the Dirac operator reduces indeed to a Dirac operator for an ADHMN construction for D2-branes ending on D4-branes. As explained in \cite{Saemann:2010cp}, this Dirac operator is a mere doubling of the one appearing in the ordinary ADHMN construction.

Correspondingly, the ultra-local part of the self-dual string equation on loop space \eqref{eq:NALoopSpaceSDS} evidently reduces to the Bogomolny equation \eqref{eq:Bogomolny}. 

In the Basu-Harvey equation, one assumes that the scalar field $X^4$ develops a vacuum expectation value in a 3-algebra direction: $\langle X^4\rangle=v$, $v\in\CA$, cf.\ \cite{Mukhi:2008ux}. To leading order in $v$, the Basu-Harvey equation then reduces to the Nahm equation \cite{Saemann:2010cp,Papageorgakis:2011xg}.

\subsection{Examples}

Let us now give some explicit examples of the above construction. The case of a single M2-brane ending on a single M5-brane corresponds to $k=N=1$ and in this case, the real 3-algebra is abelian. The Nahm data consist of constants and the Dirac operator reduces to 
\begin{equation}\label{eq:Dirack1}
 \nablabs_{s,x(\tau)}=\gamma_5\dder{s}+\tfrac{1}{2}\gamma^{\mu\nu}\oint \dd \tau \left(\di x^\mu(\tau)\xd^\nu(\tau)-X^{\mu}_0(\tau)\xd^\nu(\tau)\right)~.
\end{equation}
As above, we decompose $X^{\mu}_0(\tau)=\di x^\mu_0(\tau)\id_\CA$ and introduce the shifted loop space coordinate $y^\mu(\tau)=x^\mu(\tau)-x^\mu_0(\tau)$ as well as the modified area coordinates $y^{\mu\nu}:=\oint \dd \tau\, y^{[\mu}(\tau)\xd^{\nu]}(\tau)$. The zero modes of the Dirac operator \eqref{eq:Dirack1} are
\begin{equation}\label{eq:zmrSDk1N1}
\begin{aligned}
\psi^+_{s,x(\tau)}&\sim\de^{-r^2_- s}\left(\begin{array}{c}
\di \left(r_-^2+y^{12} -y^{34}\right) \\
y^{13}+y^{24}+\di (y^{23}-y^{14}) \\
0\\ 
0                                
\end{array}
\right)~,\\[0.2cm]
\psi^-_{s,x(\tau)}&\sim\de^{-r^2_+ s}\left(\begin{array}{c} 
0\\ 
0 \\
\di \left(r_+^2+y^{12} +y^{34}\right) \\
y^{13}-y^{24}+\di (y^{23}+y^{14})
\end{array}
\right)~,
\end{aligned}
\end{equation}
where
\begin{equation}
 r_\pm^2:=\tfrac{1}{2}\sqrt{(y^{\mu\nu}\pm\tfrac{1}{2}\eps_{\mu\nu\kappa\lambda}y^{\kappa\lambda})^2}~.
\end{equation}
The resulting Higgs field and gauge potential read as
\begin{equation}\label{eq:Phik1}
\Phi=\begin{pmatrix}
\frac{\di}{2 r_-^{2}}&0\\
0&\frac{\di}{2 r_+^{2}}                      
\end{pmatrix}
\eand
A(\sigma)=\begin{pmatrix}
A^+(\sigma)&0\\
0&A^-(\sigma)                 
\end{pmatrix}~,
\end{equation}
where 
\begin{equation}
A^+(\sigma)=\frac{\di}{2r_-^2(r_-^2+(y^{12}-y^{34}))}\left(\begin{array}{cc}
\xd^{3}(\sigma)(y^{23}-y^{14})+\xd^{4}(\sigma)(y^{13}+y^{24})\\
\xd^{4}(\sigma)(y^{23}-y^{14})-\xd^{3}(\sigma)(y^{13}+y^{24})\\
\xd^{1}(\sigma)(y^{14}-y^{23})+\xd^{2}(\sigma)(y^{13}+y^{24})\\ 
\xd^{2}(\sigma)(y^{14}-y^{23})-\xd^{1}(\sigma)(y^{13}+y^{24})                               
\end{array}
\right)~,
\end{equation}
and $A^-$ is obtained from $A^+$ by substituting $x^4(\sigma)\rightarrow -x^4(\sigma)$. Note that $A^+$ depends only on anti-self-dual combinations of area coordinates, therefore  $A^-$ depends only on self-dual combinations. Altogether, the $\au(1)_+\oplus\au(1)_-$ valued fields are functions of all six linearly independent area coordinates.

One readily checks that these fields satisfy the self-dual string equation on loop space \eqref{eq:exLoopSpaceSDS}. Note that the zero modes \eqref{eq:zmrSDk1N1} reduce to the corresponding zero modes \eqref{eq:zmMN1} in the monopole case for $k=1$, for $x^\mu(\tau)=x_0^\mu+R\delta^\mu_4\tau$ and $s\rightarrow s/r_-$, as expected.

The case $k=1$, $N=2$ has been derived with the reduced loop space derivative \eqref{eq:redloopderivative} in \cite{Papageorgakis:2011xg}. In this case, the Nahm data are trivial and the corresponding Dirac operator directly on loop space is again given by
\begin{equation}\label{eq:Dirack1N2}
 \nablabs_{s,x(\tau)}=\gamma_5\dder{s}+\tfrac{\di}{2}\gamma^{\mu\nu}\oint \dd \tau\,  x^\mu(\tau)\xd^\nu(\tau)~.
\end{equation}
Consider the interval $\CI=(-s_0,s_0)$. The zero modes of the Dirac operator \eqref{eq:Dirack1N2} on $\CI$ are
\begin{equation}
 \psi=n\left(\left(\begin{array}{cc}\cosh(r_-^2)\unit_2 & 0 \\
             0 & \cosh(r_+^2)\unit_2
            \end{array}\right)
-\frac{\di}{2}\left(\begin{array}{cc}\frac{\sinh(r_-^2)}{r_-^2}\unit_2 & 0 \\
             0 & -\frac{\sinh(r_+^2)}{r_+^2}\unit_2
            \end{array}\right)\gamma^{\mu\nu}y^{\mu\nu}\right)~,
\end{equation}
where the normalization factor $n$ reads as
\begin{equation}
 n=\left(\begin{array}{cc}\sqrt{\frac{r_-^2}{\sinh(2 s_0 r_-^2)}}\unit_2 & 0 \\
             0 & \sqrt{\frac{r_+^2}{\sinh(2 s_0 r_+^2)}}\unit_2
            \end{array}\right)~.
\end{equation}
The Higgs field resulting from formula \eqref{eq:MFieldDefinitions} is
\begin{equation}
 \Phi=\frac{\di}{2}\left(\begin{array}{cc}\frac{1}{r_-^4}\big(1-2r_-^2 s_0 \coth(2 r_-^2 s0)\big)\unit_2 & 0 \\
             0 & \frac{1}{r_+^4}\big(1-2r_+^2 s_0 \coth(2 r_+^2 s0)\big)\unit_2
            \end{array}\right)~\gamma^{\mu\nu}\gamma_5 y^{\mu\nu}~.
\end{equation}
Note that $\Phi$ takes values in the adjoint representation of $\au(2)_+\oplus\au(2)_-$. It is not clear, what gauge algebra one should expect for a pair of M5-branes. The results of \cite{Papageorgakis:2011xg}, however, suggest that this should be the associated Lie algebra of $A_4$, which is $\frg_{A_4}=\asu(2)\oplus\asu(2)$, in agreement with our result.

For the construction in the case $k=2$, $N=1$, we can use the real 3-algebra $\CCC_4$. As pointed out in the appendix, $\CCC_4$ contains $A_4$ as a sub 3-Lie algebra. We can choose the solution of the generalized Basu-Harvey \eqref{eq:r3BasuHarvey} to be 
\begin{equation}
 X^\mu=\frac{e_\mu}{\sqrt{2s}}~,
\end{equation}
where the $e_\mu$ are orthonormal generators of $A_4$ in $\CCC_4$. In the monopole case, we computed for simplicity the Higgs field at $x^3=R$. This was sufficient, as the Higgs field for $k$ coincident monopoles only depends on the radial distance. Here, we expect the Higgs field to depend only on $r_\pm^2$. It is therefore sufficient to compute the Higgs field at $y^{12}=r_-^2=r_+^2=:r^2$. Moreover, the Higgs field just depends on the ``shadow'' of the curve on the 12-plane, not its shape. We can therefore assume that the loop $x$ is a circle:
\begin{equation}
 x(\sigma)=\frac{1}{2\pi}\left(\begin{array}{c}
                  r\sin(\sigma)\\
		  r\cos(\sigma)\\
0 \\ 0
                 \end{array}\right).
\end{equation}
The zero modes of the Dirac operator \eqref{eq:r3DiracOperators} read as\footnote{In the paper \cite{Saemann:2010cp}, compatible representations of $\frg_{A_4}$ were introduced to simplify the reduction to the Nahm equation. Here, we refrain from doing this. Compatible representations could also be used for hermitian 3-algebras in the next section to give the same results.} 
\begin{equation}
\psi=\sqrt{2}r^2\sqrt{s}\de^{-r^2 s}\begin{pmatrix}e_1+\di e_2&&0\\0&&0\\0&&e_1+\di e_2\\0&&0\end{pmatrix}~.
\end{equation}
According to \eqref{eq:MFieldDefinitions}, the Higgs field reads as
\begin{equation}
 \Phi(x)=\frac{\di}{r^2}~\unit_2~,
\end{equation}
which is twice that of \eqref{eq:Phik1}. The charge is thus correctly reproduced.

In principle, we are now able to construct solutions for arbitrary $N$ and $k$ using solutions to the Basu-Harvey equation \eqref{eq:r3BasuHarvey} based on real 3-algebras. As the hermitian 3-algebras are physically more interesting, however, let us continue with these instead.

\section{Self-dual strings from hermitian 3-algebras}\label{sec:h3}

The extension of the construction of self-dual strings developed in \cite{Saemann:2010cp} to a construction involving hermitian 3-algebras is particularly interesting: Hermitian 3-algebras underlie the ABJM model, which has good chances of effectively describing stacks of multiple M2-branes. Therefore, the duality between the two effective descriptions of the configuration \eqref{diag:M2M5} from the perspective of the M2- and the M5-brane, respectively, should make use of hermitian 3-algebras.

\subsection{The Basu-Harvey equation for hermitian 3-algebras}

We start again from the configuration \eqref{diag:M2M5} of M2-branes ending on M5-branes, but we switch from a real description of this configuration to a complex one. Explicitly, we replace the four real coordinates $x^\mu$, $\mu=1,\ldots,4$, transverse to the M2-branes by two complex coordinates $z^1=x^1+\di x^2$ and $z^2=-x^3-\di x^4$. Correspondingly, the real fields $X^\mu$ appearing in the Basu-Harvey equation \eqref{eq:BasuHarvey} are replaced by two complex fields $Z^1:=X^1+\di X^2$ and $Z^2:=-X^3-\di X^4$. If we extend the range of these fields from a 3-Lie algebra to a hermitian 3-algebra, we obtain the analogue of the Basu-Harvey equation in the ABJM model. 

Recall that the BLG model has $\CN=8$ supersymmetry and correspondingly R-sym\-metry group $\sSO(8)$. In going from a real description to a complex one, we break the manifest R-symmetry group from $\sSO(8)$ to $\sSU(4)\simeq \sSO(6)$. The ABJM model is then obtained by generalizing the BLG action such that the matter fields can take values in a hermitian 3-algebra, upon which supersymmetry is indeed reduced from $\CN=8$ to $\CN=6$ in general.

Recall that the metric hermitian 3-algebra appearing in the ABJM model is $\CA={\rm Mat}_\FC(k)$ with a 3-bracket and inner product given respectively by\footnote{We use the notation $\bar{a}=a^\dagger$ as well as $\bar Z_\beta := (Z^\beta)^\dagger$ to avoid overdecorating symbols.}
\begin{equation}
 [a,b;c]:=a\bar c b-b \bar c a\eand(a,b):=\tr(\bar a b)~,~~~a,b,c\in\CA~.
\end{equation}
The metric 3-Lie algebra $A_4$ is reproduced in this way by choosing the basis
\begin{equation}
\left(\tfrac{\di}{\sqrt{2}}\sigma^1, \tfrac{\di}{\sqrt{2}}\sigma^2,\tfrac{\di}{\sqrt{2}}\sigma^3,\tfrac{1}{\sqrt{2}}\unit_2\right)~,
\end{equation}
where the $\sigma^i$, $i=1,2,3$, are the standard Pauli matrices. Using this case, we can adjust the normalization of our fields such that they match the normalization for the real 3-algebras.

The analogue of the Basu-Harvey equation in the ABJM model was previously derived in \cite{Gomis:2008vc,Terashima:2008sy,Hanaki:2008cu} and reads in our conventions as\footnote{We rescaled our fields and thus dropped the Chern-Simons level appearing in \cite{Gomis:2008vc,Terashima:2008sy,Hanaki:2008cu}.}
\begin{equation}\
\dder{s}Z^\alpha=\tfrac{1}{2}(Z^\alpha\Zb_\beta Z^\beta-Z^\beta \Zb_\beta Z^\alpha)~,~~~\alpha,\beta=1,2~.
\end{equation}
Written in the abstract 3-bracket notation explained in appendix \ref{app:hermitian3algebras}, we have
\begin{equation}\label{eq:h3Basu-Harvey}
\dder{s}Z^\alpha=\tfrac{1}{2}[Z^\alpha,Z^\beta;Z^\beta]=-\tfrac{\di}{2}D(\di Z^\beta,Z^\beta)\acton Z^\alpha~,
\end{equation}
and it is this equation that we use as a Basu-Harvey equation for hermitian 3-algebras. We inserted the factors of $\di$ in \eqref{eq:h3Basu-Harvey}, as we choose to work with antihermitian generators of $\frg_\CA$. The unusual contraction over two upper indices of $\sSU(2)$ is due to the antilinearity of the 3-bracket and the map $D(\dotsp,\dotsp)$.

\subsection{The construction}

Here we wish to rewrite the Dirac operator \eqref{eq:r3DiracOperators}  in terms of complex fields and coordinates, however to get both self-dual and anti-self-dual combinations of coordinates that appear in the lower-right and upper-left blocks, respectively, we need to introduce coordinates $\hat z_1:=z^1=x^1+\di x^2~,~~\hat z_2:=\zb^2=-x^3+\di x^4$. Now we can use 
\begin{equation}\label{eq:sigmaident}
 \gamma^{\mu\nu} x^\mu\otimes x^\nu=\tfrac{1}{4}\gamma^{\mu\nu}\left((\sigma^{\mu\nu}{}_\alpha{}^\beta (z^\alpha\otimes \zb_\beta-\zb_\beta\otimes z^\alpha)+\sigmab^{\mu\nu}{}^\alpha{}_\beta (\hat z_\alpha\otimes \hat \zb^\beta-\hat \zb^\beta\otimes\hat z_\alpha)\right)~,
\end{equation}
where we used
\begin{equation}
 \sigma^{\mu\nu}=\tfrac{1}{4}(\sigma^\mu\sigmab^\nu-\sigma^\nu\sigmab^\mu)~,~~~\sigma^\mu=(-\di\sigma^i,\unit)~,~~~\sigmab^\mu=(\di\sigma^i,\unit)~.
\end{equation}
Recall that the $\sigma^{\mu\nu}$ satisfy the identities
\begin{equation}\label{eq:identitiessigmamunu}
\begin{aligned}
{}[\sigma^{\mu\nu},\sigma^{\kappa\lambda}]&=\delta^{\nu\kappa}\sigma^{\mu\lambda}-\delta^{\mu\kappa}\sigma^{\nu\lambda}+\delta^{\mu\lambda}\sigma^{\nu\kappa}-\delta^{\nu\lambda}\sigma^{\mu\kappa}~,\\
\{\sigma^{\mu\nu},\sigma^{\kappa\lambda}\}&=\tfrac{1}{4}\left(\delta^{\nu\kappa}\delta^{\mu\lambda}-\delta^{\mu\kappa}\delta^{\nu\lambda}+\delta^{\mu\lambda}\delta^{\nu\kappa}-\delta^{\nu\lambda}\delta^{\mu\kappa}+2\eps^{\mu\nu\kappa\lambda}\right)\unit_2~,\\
\sigma^{\mu\nu}{}_\alpha{}^\beta\sigma^{\mu\nu}{}_{\gamma}{}^\delta&=\delta_\alpha^\beta\delta_\gamma^\delta-2\delta_\alpha^\delta\delta_\gamma^\beta~,~~~
\sigma^{[\mu\kappa}{}_\alpha{}^\beta\sigma^{\kappa\nu]}{}_\gamma{}^\delta=\tfrac{1}{2}(\sigma^{\mu\nu}{}_\alpha{}^\delta\delta_\beta^\gamma-\sigma^{\mu\nu}{}_\gamma{}^\beta\delta_\alpha^\delta)~.
\end{aligned}
\end{equation}
So using \eqref{eq:sigmaident} we can write the upper-left block of the Dirac operator 
\begin{equation}
 \nablas_{s,z}:=\left(\begin{array}{cc} 
\nablas^+_{s,z} & 0 \\ 
0 & \nablas^-_{s,z}
\end{array}\right)
\end{equation}
as
\begin{equation}
\begin{aligned}
 \nablas^+_{s,z}&=-\unit_2\dder{s}-\tfrac{\di}{4}\sigma^{\mu\nu}\sigma^{\mu\nu}{}_\alpha{}^\beta\left(D(\di Z^\alpha,Z^\beta)-\oint\dd \tau\,z^{\alpha}(\tau)  \dot{\zb}_\beta(\tau) - \dot z^\alpha(\tau) \bar z_\beta(\tau)\right)~,\\
 \nablabs^+_{s,z}&=+\unit_2\dder{s}-\tfrac{\di}{4}\sigma^{\mu\nu}\sigma^{\mu\nu}{}_\alpha{}^\beta\left(D(\di Z^\alpha,Z^\beta)-\oint\dd \tau\,z^{\alpha}(\tau)  \dot{\bar z}_\beta(\tau) - \dot z^\alpha(\tau) \bar z_\beta(\tau)\right)~,
\end{aligned}
\end{equation}
where $Z^\alpha\in\CA$ and $\CA$ is a metric hermitian 3-algebra. The lower-right block $\nablas^-_{s,z}$ can be written in a similar way using $\hat z_\alpha$ and $\hat Z_1:=Z^1=X^1+\di X^2~,~~\hat Z_2:=\Zb^2=-X^3+\di X^4$.

Note that as done in the real case in \eqref{eq:shiftDiracReal}, one could include an additional central part in the above Dirac operator to allow for center of mass motion of the self-dual strings. 

The first step in our construction is to verify that the Laplace operator $\Delta^+_{s,z}:=\nablabs^+_{s,z}\nablas^+_{s,z}$ is positive and central in $\sU(2)$, if the $Z^\alpha$ satisfy the Basu-Harvey equation \eqref{eq:h3Basu-Harvey}. One readily computes the non-central part of the Laplace operator to be
\begin{equation}
\sigma^{\mu\nu}\sigma^{\mu\nu}{}_\alpha{}^\beta\left(-\tfrac{\di}{2}\right)\dder{s}D(\di Z^\alpha,Z^\beta)-\tfrac{1}{4}\sigma^{\mu\nu} \sigma^{\mu\kappa}{}_\alpha{}^\beta\sigma^{\kappa\nu}{}_\gamma^\delta[D(\di Z^\alpha,Z^\beta),D(\di Z^\gamma,Z^\delta)]~.
\end{equation}
Using the fundamental identity \eqref{eq:h3fundamentalIdentity} and the identities \eqref{eq:identitiessigmamunu} simplifies this further to
\begin{multline}
\sigma^{\mu\nu}\sigma^{\mu\nu}{}_\alpha{}^\beta\tfrac{1}{2}\dder{s}D(Z^\alpha,Z^\beta)\\+\tfrac{1}{8}\sigma^{\mu\nu}(\sigma^{\mu\nu}{}_\alpha{}^\delta\delta_\beta^\gamma-\sigma^{\mu\nu}{}_\gamma{}^\beta\delta_\alpha^\delta)\big(D([Z^\gamma,Z^\alpha;Z^\beta],Z^\delta)-D(Z^\gamma,[Z^\delta,Z^\beta;Z^\alpha])\big)~.
\end{multline}
Due to $\sigma^{\mu\nu}{}_\alpha{}^\delta\eps^{\beta\alpha}=\sigma^{\mu\nu}{}_\alpha{}^\beta\eps^{\delta\alpha}$, we have
\begin{equation}
\begin{aligned}
 -\sigma^{\mu\nu}{}_\gamma{}^\beta D([Z^\gamma,Z^\alpha;Z^\beta],Z^\alpha)&=\sigma^{\mu\nu}{}_\alpha{}^\delta D([Z^\beta,Z^\alpha;Z^\beta],Z^\delta)~,\\
 -\sigma^{\mu\nu}{}_\alpha{}^\delta D(Z^\beta,[Z^\delta,Z^\beta;Z^\alpha])&=\sigma^{\mu\nu}{}_\gamma{}^\beta D(Z^\gamma,[Z^\alpha,Z^\beta;Z^\alpha])~,
\end{aligned}
\end{equation}
and the non-central part of the Laplace operator becomes proportional to
\begin{equation}
 \sigma^{\mu\nu}\sigma^{\mu\nu}{}_\alpha{}^\beta\left(\dder{s}D(Z^\alpha,Z^\beta)+\tfrac{1}{2}D([Z^\gamma,Z^\alpha;Z^\gamma],Z^\beta)+\tfrac{1}{2}D(Z^\alpha,[Z^\gamma,Z^\beta;Z^\gamma]\right)~.
\end{equation}
This expression vanishes, if the Basu-Harvey equation \eqref{eq:h3Basu-Harvey} is satisfied. In this case, the Laplace operator $\Delta^-_{s,z}:=\nablabs^-_{s,z}\nablas^-_{s,z}$ and thus $\Delta_{s,z}:=\nablabs_{s,z}\nablas_{s,z}$ are positive and central in $\sU(2)$, too. Note that the inverse statement is not necessarily true, as the map $D:\CA\times\CA\rightarrow \frg_\CA$ could be degenerate. 

As in the case of real 3-algebras, we again have $2N$ zero modes $\psi_{s,z}^a\in W^{0,2}(\CI)\otimes \FC^2\otimes\FC^N\otimes \CA$, $a=1,\ldots,2N$, of the Dirac operator $\nablabs_{s,z}$. We sort them according to their chirality and normalize them such that
\begin{equation}\label{eq:abjmnormalizePsiM}
 \delta^{ab}=\tr\int_\CI\dd s\, (\psib^{a}_{s,z},\psi^{b}_{s,z})~,
\end{equation}
where $(\dotsp,\dotsp)$ denotes the inner product on $\FC^4\otimes \CA$. Contrary to the real case, we now define a complex gauge potential,
\begin{subequations}\label{eq:h3FieldDefinitions}
\begin{equation}
\begin{aligned}
\big(A_{(\alpha\tau)}\big)^{ab}&=\int \dd s\, \left(\psib^a_{s,z}, \delder{z^\alpha(\tau)}\psi^b_{s,z}\right)~,\\
\big(A^{(\alphab\tau)}\big)^{ab}&=\int \dd s\, \left(\psib^a_{s,z}, \delder{\zb_\alpha(\tau)}\psi^b_{s,z}\right)~,
\end{aligned}
\end{equation}
and a scalar field
\begin{equation}
 \Phi^{ab}=\di \int \dd s\, \left(\psib^a_{s,z}\,,s\,\psi^b_{s,z}\right)~.
\end{equation}
\end{subequations}
These fields take values in the gauge algebra $\au(N)_+\oplus\au(N)_-$. The self-dual string equation on loop space \eqref{eq:NALoopSpaceSDS} for the $\au(N)_+$-components of the complex gauge potential and the Higgs field reads as
\begin{equation}\label{eq:h3NASDSeq}
\begin{aligned}
 F_{(\alpha\sigma)(\beta\tau)}=[\nabla_{(\alpha\sigma)},\nabla_{(\beta\tau)}]&=\tfrac{1}{2}(\dot{\zb}_{\beta}(\sigma)\nabla_{(\alpha\tau)}\Phi-\dot{\zb}_{\alpha}(\tau)\nabla_{(\beta\sigma)}\Phi)~,\\
 F^{(\alphab\sigma)(\betab\tau)}=[\nabla^{(\alphab\sigma},\nabla^{(\betab\tau)}]&=\tfrac{1}{2}(\dot{z}^\beta(\sigma)\nabla^{(\alphab\tau)}\Phi-\dot{z}^\alpha(\tau)\nabla^{(\betab\sigma)}\Phi)~,\\
 F_{(\alpha\sigma)}{}^{(\betab\tau)}=[\nabla_{(\alpha\sigma)},\nabla^{(\betab\tau)}]&=\tfrac{1}{2}\eps_{\alpha\gamma}\eps^{\beta\delta}(\dot{z}^\gamma(\tau) \nabla_{(\delta\sigma)}\Phi-\dot{\zb}_\delta(\sigma)\nabla^{(\gammab\tau)}\Phi)~,
\end{aligned}
\end{equation}
where $\nabla_{(\alpha\sigma)}:=\delder{z^\alpha(\sigma)}+A_{(\alpha\sigma)}$, $\nabla^{(\alphab\sigma)}:=\delder{\zb_\alpha(\sigma)}+A^{(\alphab\sigma)}$ and $\eps^{12}=-\eps_{12}:=1$. The corresponding equations for the $\au(N)_-$ components are obtained from \eqref{eq:h3NASDSeq} by substituting $z\rightarrow\hat z$.

The proof that the fields \eqref{eq:h3FieldDefinitions} indeed satisfy these equations closely follows the real case. For simplicity, we restrict to the $\au(N)_+$ components. The proof for the $\au(N)_-$ components is completely analogous. We start by introducing the Green's function $G_{z}(s,t)$ of the Laplace operator $\Delta^+_{s,z}$ leading again to the completeness relation
\begin{equation}\label{eq:abjmcompleteness}
 \delta(s-t)=\psi^a_{s,z}\big(\psib^a_{t,z},\dotsp\big)-\nablas^+_{s,z} G_z(s,t)\nablabs^+_{t,z}~.
\end{equation}
We then compute
\begin{equation}
\begin{aligned}
(F_{(\alpha\sigma)(\beta\tau)})^{ab}&=2\int_\CI \dd s\,(\delta_{[(\alpha\sigma)}\psib^a_{s,z},\delta_{(\beta\tau)]}\psi^b_{s,z})+2\!\int_\CI \dd s\!\int_\CI \dd t\,(\psib^a_{s,z},\delta_{[(\alpha\sigma)}\psi^c_{s,z})(\psib^c_{t,z},\delta_{(\beta\tau)]}\psi^b_{t,z})\\
&=-2\int_\CI \dd s\int_\CI \dd t\,\Big(\delta_{[(\alpha\sigma)}\psib^a_{s,z}\,,\,\left(\nablas^+_{s,z}G_z(s,t)\nablabs^+_{t,z}\right)\delta_{(\beta \tau)]}\psi^b_{t,z}\Big)
\end{aligned}
\end{equation}
and
\begin{equation}
(F_{(\alpha\sigma)}{}^{(\betab\tau)})^{ab}=-2\int_\CI \dd s\int_\CI \dd t\,\Big(\delta_{[(\alpha\sigma)}\psib^a_{s,z}\,,\,\left(\nablas^+_{s,z}G_z(s,t)\nablabs^+_{t,z}\right)\delta^{(\betab \tau)]}\psi^b_{t,z}\Big)~.
\end{equation}
Here, we need the identities
\begin{equation}
\begin{aligned}
 \sigma^{\mu\nu}\sigma^{\kappa\lambda}\big(\sigma^{\mu\nu}{}_{\alpha}{}^\gamma\sigma^{\kappa\lambda}{}_{\beta}{}^\delta\dot{\zb}_\gamma(\sigma)\dot{\zb}_\delta(\tau)\big)\stackrel{[\cdot]}{=}
2\sigma^{\mu\nu}\sigma^{\mu\nu}{}_{\alpha}{}^\gamma\dot{\zb}_\gamma(\tau)\dot{\zb}_{\beta}(\sigma)~,\hspace{2cm}\\
\sigma^{\mu\nu}\sigma^{\kappa\lambda}\big(\sigma^{\mu\nu}{}_\alpha{}^\gamma\sigma^{\kappa\lambda}{}_\delta{}^\beta \dot{\zb}_\gamma(\sigma)\dot{z}^\delta(\tau)-\sigma^{\mu\nu}{}_\delta{}^\beta\sigma^{\kappa\lambda}{}_\alpha{}^\gamma\dot{z}^\delta(\tau)\dot{\zb}_\gamma(\sigma)\big)\hspace{4cm}\\
=-2\epsilon_{\alpha \gamma}\epsilon^{\beta\delta}\sigma^{\mu\nu}(\sigma^{\mu\nu}{}_\kappa{}^\gamma \dot z^\kappa(\tau)\dot{\zb}_\delta(\sigma)+\sigma^{\mu\nu}{}_\delta{}^\kappa \dot z^\gamma(\tau)\dot{\zb}_\kappa(\sigma))~,
\end{aligned}
\end{equation}
where $\stackrel{[\cdot]}{=}$ denotes weighted antisymmetrization under $(\alpha\sigma)\leftrightarrow (\beta\tau)$. The identities lead to
\begin{equation}
\begin{aligned}
(F_{(\alpha\sigma)(\beta\tau)})^{ab}&\stackrel{[\cdot]}{=}\int_\CI \dd s\int_\CI \dd t\,\Big(\psib_{s,z}^a,\,\left(\sigma^{\mu\nu}\sigma^{\mu\nu}{}_{\alpha}{}^\gamma\dot{\zb}_{(\gamma\tau)}\dot{\zb}_{(\beta\sigma)} G_z(s,t)\right)\psi^b_{t,z}\Big)\\
&=\di\dot{\zb}_{[(\beta\sigma)}\int_\CI \dd s\,\Big(\nabla_{(\alpha\tau)]}\psib^a_{s,z},\,s\,\psi^b_{s,z}\Big)+\Big(\psib^a_{s,z},\,s\,\nabla_{(\alpha\tau)]}\psi^b_{s,z}\Big)\\
&=\tfrac{1}{2}(\dot{\zb}_{\beta}(\sigma)\nabla_{(\alpha\tau)}\Phi^{ab}-\dot{\zb}_{\alpha}(\tau)\nabla_{(\beta\sigma)}\Phi^{ab})~,
\end{aligned}
\end{equation}
and
\begin{equation}
 F_{(\alpha\sigma)}{}^{(\betab\tau)}=\tfrac{1}{2}\eps_{\alpha\gamma}\eps^{\beta\delta}(\dot{z}^\gamma(\tau) \nabla_{(\delta\sigma)}\Phi-\dot{\zb}_\delta(\sigma)\nabla^{(\gammab\tau)}\Phi)~.
\end{equation}

\subsection{Comment on the reduction to monopoles}

In the complex description of self-dual strings we work with loops wrapping the $x^4$-direction by imposing the condition $\dot{\zb}_\alpha=-\di R\delta_\alpha^2$, cf.\ section \ref{sec:r3reduction}. Then the whole reduction procedure for hermitian 3-algebras works fully analogously to the case of real 3-algebras. We therefore refrain from going into further details.

\subsection{Examples}

We now present a few simple examples of our construction. We start with the simplest case $k=N=1$, which is a mere rewriting of the same case for real 3-algebras in complex notation. We can rewrite $r_{-}^2=\sqrt{\frac{1}{4}z^\alpha{}_\alpha z^\beta{}_\beta-\frac{1}{2}z^\alpha{}_\beta z^\beta{}_\alpha}$, where we've used complex area coordinates: $z^\alpha{}_\beta:=\frac{1}{2}\int\dd\tau(z^{\alpha}(\tau)\dot{\zb}_\beta(\tau)-\dot z^\alpha(\tau)\zb_\beta(\tau))$. As in the real case, the Nahm data are trivial: $Z^\alpha=0$ and the zero mode reads before normalization as
\begin{equation}
\psi^+\sim\de^{-r_{-}^2 s}\begin{pmatrix} \di r_{-}^2 +z^1{}_1-{z}^2{}_2\\ 2z^1{}_2\\ 0 \\ 0\end{pmatrix}~,~~~
\psi^-\sim\de^{-r_{+}^2 s}\begin{pmatrix} 0 \\ 0 \\ \di r_{+}^2 +\hat{z}^1{}_1-\hat{z}^2{}_2\\ 2\hat{z}^1{}_2\end{pmatrix}~,
\end{equation}
and leads to the expected Higgs field
\begin{equation}
\Phi=\left(\begin{array}{cc}
\frac{\di}{2r_{-}^2} & 0 \\ 0 &
\frac{\di}{2r_{+}^2}
           \end{array}\right)~.
\end{equation}

Next, let us consider the case $N=1$, $k$ arbitrary. Note that for $k>2$, this case could not have been treated using 3-Lie algebras. The corresponding solution to the Basu-Harvey equation has been found in \cite{Gomis:2008vc}. In our conventions, it reads as
\begin{equation*}
 Z^1=\frac{1}{\sqrt{s}}\left(\begin{array}{ccccc} 
	    0 &  0 & 0 & \ldots & 0 \\ 
	    0 & \sqrt{1} & 0 & & \vdots \\ 
	    0  & 0 & \sqrt{2} & & \\
	    \vdots  & & & \ddots & \\
	    0 & \ldots & & & \sqrt{k-1}
	  \end{array}\right),~
 Z^2=\frac{1}{\sqrt{s}}\left(\begin{array}{ccccc} 
	    0 &  0 & 0 & \ldots & 0 \\ 
	    \sqrt{k-1} & 0 & 0 & & \vdots \\ 
	    0  & \sqrt{k-2} & 0 & & \\
	    \vdots  & & \ddots &  & \\
	    0 & \ldots & 0 & 1 & 0
	  \end{array}\right),
\end{equation*}
As before, we consider the zero modes only at $y^{41}=r_{\pm}^2=\di z^1{}_2=\di z^2{}_1=:r$ and extract the Higgs field as a consistency check. The zero modes of the Dirac operator $\nablabs_{s,z}$ with this restriction are given by 
\begin{equation*}
 \psi^+\sim\de^{-r^2 s}s^{\frac{k-1}{2}}\left(\begin{array}{c}\zeta\\ \zeta\\ 0 \\ 0\end{array}\right)\eand 
 \psi^-\sim\de^{-r^2 s}s^{\frac{k-1}{2}}\left(\begin{array}{c}0 \\ 0\\ \zeta\\ \zeta\end{array}\right)
\end{equation*}
with
\begin{equation}
 \zeta=\left(\begin{array}{ccccc}\sqrt{\binomr{k-1}{0}} & \sqrt{\binomr{k-1}{1}}\ & \sqrt{\binomr{k-1}{2}}\ & \ldots & \sqrt{\binomr{k-1}{k-1}}\\ 0 & 0 & 0 & 0 & 0\\
             &&\vdots&&\\
0 & 0 & 0 & 0 & 0
\end{array}\right)~.
\end{equation}
One readily computes the Higgs field 
\begin{equation}
 \Phi=\frac{\di k}{2r^2}\,\unit_2~.
\end{equation}
and we indeed recovered a self-dual string of charge $k$.

\section*{Acknowledgments}

CS would like to thank Martin Wolf for helpful discussions. This work was supported by a Career Acceleration Fellowship from the UK Engineering and Physical Sciences Research Council.

\appendices

The 3-algebraic structures first used in the BLG model are so-called 3-Lie algebras \cite{Filippov:1985aa}. As they turned out to be too rigid for an effective description of stacks of arbitrarily many M2-branes, various generalizations have been proposed. We are interested in the so-called\footnote{The algebras we define in the following are also known as generalized 3-Lie algebras and hermitian 3-Lie algebras.} {\em real 3-algebras} \cite{Cherkis:2008qr} as well as the {\em hermitian 3-algebras} \cite{Bagger:2008se}, see also \cite{deMedeiros:2008zh} for further details.

\subsection{Real 3-algebras}\label{app:real3algebras}

A {\em real 3-algebra} is a real vector space $\CA$ endowed with a 3-bracket $[\dotsp,\dotsp,\dotsp]:\CA^{\wedge 2}\times\CA \rightarrow \CA$, which is trilinear and antisymmetric in its first two slots. Moreover, it satisfies the {\em fundamental identity}
\begin{equation}\label{eq:fundamentalIdentity}
 [a,b,[c,d,e]]=[[a,b,c],d,e]+[c,[a,b,d],e]+[c,d,[a,b,e]]
\end{equation}
for all $a,b,c,d,e\in\CA$. If we endow $\CA$ with a metric $(\dotsp,\dotsp):\CA\odot\CA\rightarrow \FR$, which satisfies the {\em compatibility condition}
\begin{equation}
 ([a,b,c],d)+(c,[a,b,d])=0~,
\end{equation}
we arrive at a {\em metric real 3-algebra}. This notion extends that of a {\em metric 3-Lie algebra}, for which the 3-bracket is totally antisymmetric.

A real 3-algebra $\CA$ comes with an associated Lie algebra $\frg_\CA$ of inner derivations. The vector space of inner derivations is the linear span of $D(a,b)$, $a,b\in\CA$, where 
\begin{equation}
 D(a,b)\acton c:=[a,b,c]~,~~~c\in\CA~.
\end{equation}
The commutator of two inner derivations is again an inner derivation due to the fundamental identity \eqref{eq:fundamentalIdentity}.

As an example, consider the family of metric real 3-algebras $\CCC_{2d}$ \cite{Cherkis:2008qr}. The underlying vector space of $\CCC_{2d}$ is spanned by hermitian matrices of dimension $2d\times 2d$ which anticommute with $\Gamma_{\rm ch}=\diag(\unit_d,-\unit_d)$. The 3-bracket is given by
\begin{equation}
 [a,b,c]:=\tfrac{1}{4}[[a,b]\Gamma_{\rm ch},c]~,~~~a,b,c\in\CCC_{2d}~,
\end{equation}
and together with the scalar product
\begin{equation}
 (a,b):=\tr(a b)~,~~a,b\in\CCC_{2d}~,
\end{equation}
this is a metric real 3-algebra with associated Lie algebra $\frg_{\CCC_{2d}}=\asu(d)\oplus \asu(d)$. Note that we can embed the Weyl representation of the generators $\gamma^\mu$ of the Clifford algebra $C\ell(\FR^4)$ into $\CCC_{4}$. Restricting to the vector space spanned by the $\gamma^\mu$, we recover the 3-Lie algebra $A_4$ with 3-bracket \cite{Gustavsson:2007vu}
\begin{equation}
 [\gamma^\mu,\gamma^\nu,\gamma^\rho]:=\tfrac{1}{4}[[\gamma^\mu,\gamma^\nu]\gamma_5,\gamma^\rho]=\eps^{\mu\nu\rho\sigma}\gamma^\sigma~,
\end{equation}
where $\gamma_5=\gamma^1\gamma^2\gamma^3\gamma^4$. The associated Lie algebra of the 3-Lie algebra $A_4$ is $\frg_{A_4}=\asu(2)\oplus\asu(2)$.

\subsection{Hermitian 3-algebras}\label{app:hermitian3algebras}

A {\em hermitian 3-algebra} is a complex vector space $\CA$ endowed with a 3-bracket $[\dotsp,\dotsp;\dotsp]$, which is linear and antisymmetric in its first two slots and antilinear in its third slot. Furthermore, it satisfies the following {\em fundamental identity} 
\begin{equation}\label{eq:h3fundamentalIdentity}
 [[a,b;c],d;e]=[[a,d;e],b;c]+[a,[b,d;e];c]-[a,b;[c,e;d]]
\end{equation}
for all $a,b,c,d,e\in\CA$. Adding a hermitian form $(\dotsp,\dotsp)$ on $\CA$ which satisfies the {\em compatibility condition}
\begin{equation}
([a,b;c],d)=(b,[c,d;a])
\end{equation}
turns $\CA$ into a {\em metric hermitian 3-algebra}.

Analogously to a real 3-algebra, a hermitian 3-algebra $\CA$ comes with a Lie algebra of inner derivations $\frg_\CA$. As a vector space, these are spanned by 
\begin{equation}
D(a,b)\acton c := [c,a;b]~, 
\end{equation}
and the Lie bracket $[X,Y]$, $X,Y\in\frg_\CA$, closes due to the fundamental identity \eqref{eq:h3fundamentalIdentity}. Note that the map $D(\dotsp,\dotsp)$ is linear in its first slot and antilinear in its second slot. Moreover, note that $\frg_\CA$ is a complex Lie algebra. A real subalgebra $\frg_\CA^R$ can be constructed from \cite{deMedeiros:2008zh}
\begin{equation}
E(a,b):=\tfrac{1}{2}(D(a,b)-D(b,a))~.
\end{equation}

The most important example of hermitian 3-algebras are those appearing in the ABJM model. That is, we identify $\CA$ with complex matrices of dimension $d\times d$ together with the 3-bracket
\begin{equation}
[c,a;b] :=a \bar b c- c \bar b a
\end{equation}
and the hermitian form 
\begin{equation}
 (a,b):= \tr(\bar{a} b)~.
\end{equation}
The associated Lie algebra $\frg_\CA$ is $\au_f(d)\oplus \au_{\bar{f}}(d)$ and $\CA$ forms the fundamental representation under $\au_f(d)$ and the antifundamental representation under $\au_{\bar{f}}(d)$.

\subsection{Jacobi elliptic functions and generalizations}\label{app:Jacobi}

An elliptic function is a doubly-periodic, meromorphic\footnote{Note that any doubly-periodic, holomorphic function must be constant.} function and any such function can be expressed in terms of {\em Jacobi} (or {\em Weierstra\ss}) {\em elliptic functions}. The Jacobi functions satisfy the relations\footnote{Many more relations can be found at \href{http://functions.wolfram.com}{functions.wolfram.com}.} 
\begin{equation}\label{C1}
\begin{aligned}
{\rm sn}_0z=\sin z\ ,\ \ {\rm cn}_0z=\cos z\ ,\ \ {\rm dn}_0z=1\ ,\  \ &{\rm cn}_k^2z+{\rm sn}_k^2z=1\ ,\ \ {\rm dn}_k^2z+k^2{\rm sn}_k^2z=1~,\\
{\rm sn}_kz={\rm sn}_k(z+4K(k))={\rm sn}_k(z+2\di K(k'))=-&{\rm sn}_k(z+2K(k))=\frac{{\rm sn}_{k^{-1}}kz}{k}=\frac{-\di {\rm sn}_{k'}\di z}{{\rm cn}_{k'}\di z}~,\\
{\rm cn}_k0={\rm dn}_k0=1\ ,\ \ {\rm sn}_k0=0\ ,\ \ {\rm sn}_k(z+K(k))&=\frac{{\rm cn}_kz}{{\rm dn}_kz}\  ,\ \ \ {\rm sn}_k(z+\di K(k'))=\frac{1}{k\ {\rm sn}_kz}~,\\
{\rm cn}_k({\rm sn}_k^{-1}s)=\sqrt{1-s^2}\ ,\ \ {\rm dn}_k({\rm sn}_k^{-1}s)=&\sqrt{1-k^2s^2}\ ,\ \ \dder{s}{\rm sn}_ks={\rm cn}_ks\ {\rm dn}_ks~,
\end{aligned}
\end{equation}
where $K(k)={\rm sn}_k^{-1}(1)$ and $k'^2=1-k^2$. 

They can be defined in terms of theta functions (which are not doubly-periodic) or in terms of integrals. Since the Jacobi functions are related, it suffices to define
\begin{equation}
{\rm sn}_k^{-1}(s)=\int_0^s\frac{\dd t}{\sqrt{(1-t^2)(1-k^2t^2)}}~.
\end{equation}

A {\em generalized Jacobi elliptic function} \cite{Pawellek:2009er} is given by 
\begin{equation}\label{genjacdefinition}
S^{-1}(s,k_1,k_2)=\int_0^s\frac{\dd t}{\sqrt{(1-t^2)(1-k_1^2t^2)(1-k^2_2t^2)}}~.
\end{equation}
The function $S(s,k_1,k_2)$ is hyperelliptic but can be viewed as a single-valued meromorphic function on a Riemann surface of genus two \cite{Pawellek:2009er}. It has been shown to be related to the Jacobi elliptic functions by
\begin{equation}\label{genjac}
S(s,k_1,k_2)=\frac{{\rm sn}_{\kappa}(k'_2s)}{\sqrt{k'^2_2+k_2^2{\rm sn}^2_{\kappa}(k'_2s)}}~,
\end{equation}
where $\kappa^2=\frac{k_1^2-k_2^2}{1-k_2^2}$ and $k'^2_2=1-k_2^2$.

\end{document}